\documentclass[12pt]{spieman}  
\usepackage{amsmath,amsfonts,amssymb}
\usepackage{graphicx}
\usepackage{setspace}
\usepackage{hyperref}
\usepackage{tocloft}
\usepackage{rotating}

\makeatletter
\let\c@lofdepth\relax
\let\c@lotdepth\relax
\makeatother
\usepackage{subfigure}

\usepackage[T1]{fontenc}

\usepackage{listings}
\usepackage{xcolor}

\lstdefinelanguage{yaml}{
  sensitive=false,
  keywords={true,false,null,y,n},
  keywordstyle=\bfseries, 
  comment=[l]{\#},
  commentstyle=\itshape\color{gray}, 
  stringstyle=\ttfamily,
  literate={:}{{{\bfseries{:}}}}{1} 
           {-}{{{\bfseries{-}}}}{1},
  breaklines=true,
}

\makeatletter
\def\ext@listing{lol} 
\makeatother

\usepackage{caption}
\DeclareCaptionType{listing}[Listing][List of Listings]

\usepackage{multirow}

\usepackage{lineno}

\usepackage{soul}
\usepackage{xcolor}

\title{Introducing a new generation Adaptive Optics simulation framework: from PASSATA to SPECULA }

\author[a,b,*]{Fabio Rossi}
\author[a,b,*]{Alfio Puglisi}
\author[a,b,*]{Guido Agapito}
\affil[a]{INAF -- Osservatorio Astrofisico di Arcetri, Largo E. Fermi 5, 50125, Firenze, Italy}
\affil[b]{ADaptive Optics National laboratory in Italy (ADONI)}

\cftpagenumbersoff{figure}
\cftpagenumbersoff{table} 
\begin{document} 
\maketitle

\begin{abstract}
Numerical end-to-end simulation in Adaptive Optics (AO) is a key tool in the development of complex systems, from the initial design to the commissioning phase. 
Based on our previous experience with PASSATA, we decided to develop a new AO simulation framework in Python language, 
naming it SPECULA (short for: \textbf{S}calable \textbf{P}arallel \textbf{E}xecution of \textbf{C}omputations \textbf{U}pscaling \textbf{L}arge \textbf{A}daptive optics simulations). Following an object-oriented approach, the physical entities are modeled as processing objects connected to each other to exchange data objects. A simulation is run by providing its description instead of writing and executing a specific script.
The Python language and its library flexibility allowed us to write one single code that can be run on CPU and GPU platforms. We put a strong focus on computational efficiency, relying on CuPy and its interface to access the CUDA-stream mechanism.  Moreover, SPECULA is capable of distributed computations over multiple processing nodes, making it suitable to run in an HPC environment, as tested on the Italian supercomputer Leonardo.
SPECULA can also be used in laboratory environment to implement a hybrid simulation, allowing us to interface simulated and concrete objects: this feature was demonstrated in the Adaptive Optics laboratories at Arcetri Observatory.
In this paper, we describe the main characteristics of SPECULA, show some relevant examples of its use, and finally draw our goals for the future.

\end{abstract}


{\noindent \footnotesize\textbf{*}Send all correspondence to \linkable{fabio.rossi@inaf.it}, \linkable{alfio.puglisi@inaf.it} and \linkable{guido.agapito@inaf.it}}

\begin{spacing}{1}   

\section{Introduction}
\label{sect:intro}  

Arcetri Observatory has a long experience with Adaptive Optics simulation tools: in the 2000s the first simulations of a Pyramid\cite{1996JMOp...43..289R} Wavefront Sensor (WFS)  were performed and in the 2010s the development of PASSATA\cite{2016SPIE.9909E..7EA} was crucial to support the design of the FLAO system\cite{2010SPIE.7736E..09E} for LBT\cite{2012SPIE.8444E..1AH}.
PASSATA has been continuously improved and extended to simulate different AO systems such as SOUL\cite{2023aoel.confE..80P}, GMT NGAO\cite{2014SPIE.9148E..2MP}, ERIS\cite{2023A&A...674A.207D}, MAVIS\cite{2021Msngr.185....7R}, MORFEO\cite{2024SPIE13097E..22C}, ANDES\cite{2024SPIE13097E..4WP}. Despite the good success of PASSATA as a simulation tool and its track record of proven applications, it presents a few limitations. Its implementation based on the IDL language makes the lab use difficult due to the lack of interface libraries, it allows to implement GPU code only by direct CUDA\cite{NVIDIA2025} programming and makes the scale up to HPC environment complicated. Moreover, IDL scripting is required to specify and run simulations, and many years of incremental development resulted in a heavy codebase. We also believe that its implementation language prevented it from being more widely adopted.

Considering that in the last decade Python has become the programming language of choice in astronomy, from data pipelines to instrument control, its choice for the implementation of a new simulation framework is quite obvious. It can be argued that Python is much less efficient than other programming languages when it comes to performance; however, the availability of optimized numerical libraries bridges most of this performance gap. Although we focused on performance, we have to remember that the realm in which we are in is not a hard real-time environment: our priority is to build a framework that is relatively easy to use, inspect, understand, and expand for new contributors.

We adopted the concepts and patterns used in PASSATA that proved to work well, trying to redesign what proved to be weak. So, also in SPECULA, the structural and temporal handling is taken care by a few housekeeping (or management) objects, while the simulated entities and the exchanged data are modeled, respectively, as processing and data objects. Several improvements are introduced: in particular, the automatization of object construction, which is based only on the information contained in the simulation description files (written in yaml format), and the avoidance of GPU-specific code, which is made possible by the platform-agnostic NumPy\cite{harris2020array}/CuPy\cite{okuta2017numpycompatible} code.
Similarly to what was possible in PASSATA, the user can allocate every object on the GPU or on the CPU, but this mechanism is extended in SPECULA to be able to allocate a processing object to a GPU of a specific index, when multiple GPUs are available, and to a specific process in the case of a multi-process computation, which is an essential feature in HPC environment. The access to these powerful computational resources is particularly useful in the simulation of complex and large AO systems, such as MORFEO, where the memory requirements of the simulation can easily exceed the available memory capacity of a single GPU.

In the AO community, many other AO simulators have been developed, a not exhaustive list includes: CAOS\cite{2005MNRAS.356.1263C}, CEO\cite{2023aoel.confE..56R}, COMPASS\cite{2014SPIE.9148E..6OG}, DASP\cite{2018SoftX...7...63B}, MAOS\cite{Wang2012FastEM}, OCTOPUS \footnote{\href{www.eso.org/sci/facilities/develop/ao/tecno/octopus.html}{www.eso.org/sci/facilities/develop/ao/tecno/octopus.html}}, OOMAO\cite{2014SPIE.9148E..6CC}, OOPAO\cite{2023aoel.confE..50T}, SOAPY\cite{2016SPIE.9909E..7FR}, XAOSIM \footnote{\href{github.com/fmartinache/xaosim}{github.com/fmartinache/xaosim}}, YAO\cite{2013aoel.confE..18R}. Comparison and cross-check of the simulation results with these libraries is beyond the scope of this paper; however, we welcome the authors and users of these frameworks to consider SPECULA and to contact us for comments, suggestions, and possible contributions. SPECULA is an open source project; its complete codebase is available at:

\href{github.com/ArcetriAdaptiveOptics/SPECULA}{github.com/ArcetriAdaptiveOptics/SPECULA},
while more details, examples, and tutorials can be found in the documentation, available at:
\href{specula.readthedocs.io/en/latest/}{specula.readthedocs.io}.

The remainder of this document is organized as follows.
We present the structure of the framework in Sect.\ref{sec:struct} and how it can be configured to perform a simulation in Sect.\ref{sec:conf}.
The details on how to store the simulation results on disk are described in Sect.\ref{sec:store}, while the available tools for data display are presented in Sect.\ref{sec:visual}.
We go into the details of the multiprocessing support in Sect.\ref{sec:mpi}; in Sects.\ref{sec:hwloop} and \ref{sec:compare} we report, respectively, about the interaction with physical devices and about the comparison of simulation results with those obtained in PASSATA.
Finally, in Sect.\ref{sec:conclusion}, we draw our conclusions and goals for the future.

\section{The framework}\label{sec:struct}

\subsection{Basic Concepts}\label{sec:basicconcepts}

A simulation in SPECULA is a network of \emph{processing objects} whose inputs and outputs can be connected in various ways, to update and exchange \emph{data objects}. Some specific \emph{management objects} (also now in PASSATA as \emph{houskeeping objects}) take care of the core tasks needed to run the simulation. \emph{Processing} objects represent devices, physical entities or processes (like atmosphere evolution and propagation, wavefront sensors, slope computers, deformable mirrors), while \emph{data} objects wrap the data relative to physical quantities and provide methods to access them (for example: electric field,  camera pixels, deformable mirror commands).
\emph{Management} objects implement the functionalities needed to build and run the simulation, manage data saving, and perform basic utility tasks.
Processing objects exchange information by the connection of their inputs and outputs, and do not have other references to each other, while they can reference constant data objects of common use.

The simulation network topology is static, which means that the connections between objects do not change during its temporal evolution.
The time progress is discrete, advancing by a user-configurable, fixed, and constant time step, with the only requirement for correct operation being that all internal timings (e.g. detector integration times) of processing objects are multiples of its duration.
Once the simulation process begins, at each time step, all the processing objects can be executed in any order that does not violate the partial-order relation given by the network connections: this means that there are precedences between the different objects' execution.

\begin{figure}[t]
    \begin{center}
        \includegraphics[trim={0 3cm 0 3cm},scale=0.5]{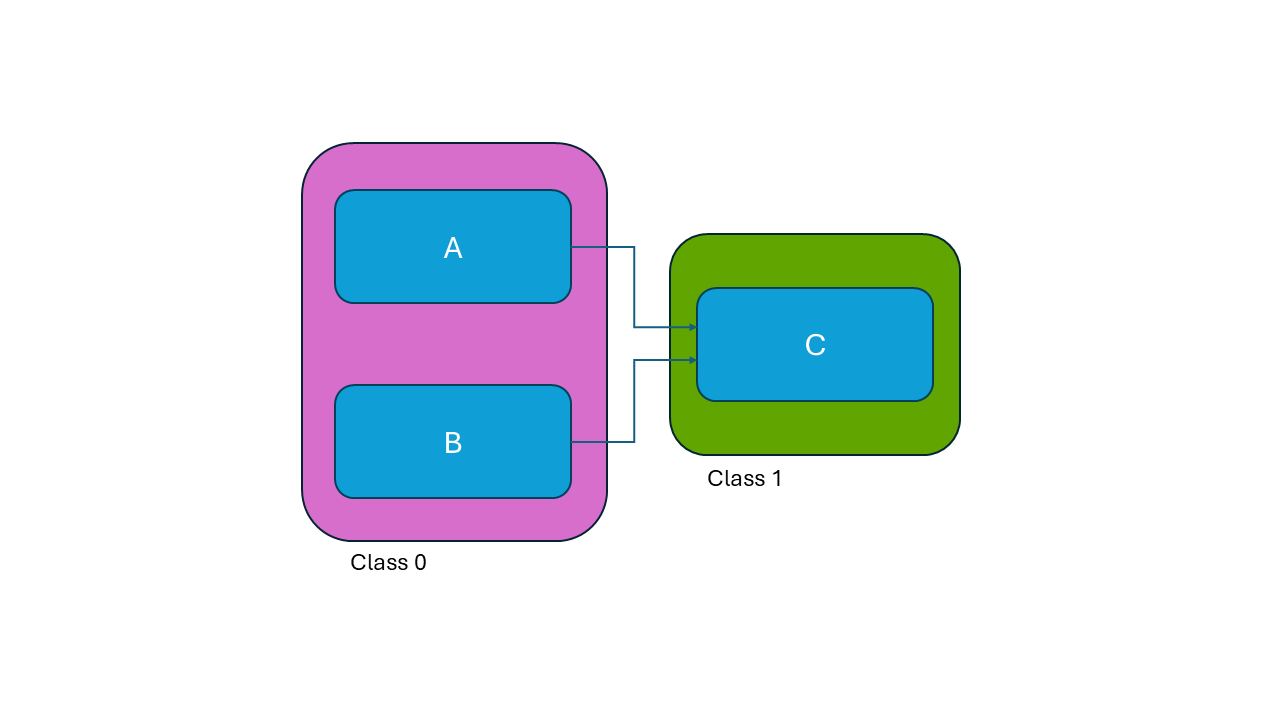}
        \caption{The nodes A and B are in the same class, both are direct predecessors of C.}
        \label{fig:partailorder}
    \end{center}
\end{figure}

Looking at the example in Fig.\ref{fig:partailorder}, since one output of A servers as input for C, A must be executed before C; similarly, B should also be executed before C, while there are no constraints between A and B: they are in the same class and, in our terminology, they will be assigned the same trigger order (an integer value).

The data objects carry the temporal information of their last modification by a processing object. At each time step, a scheduler executes all processing objects, respecting their partial-order relation. The execution of a processing object includes various phases: gathering the inputs, performing some computations, and updating its outputs.

In Adaptive Optics we usually have at least one loop in the block diagram of our system: typically the electric field which bounces off a deformable mirror is fed back to the propagation computation as one of its inputs. This particular type of  connection is associated to a delay of one time step (we call it ``delayed connection''), so it can be temporarily removed from the simulation graph and handled separately at the beginning of the next time step.
A processing object can internally buffer its inputs and have its outputs computed as a function of n previous time steps, so this includes the implementation of delays of any integer or fractional number of time steps, independently of the fact that its outputs are sent into a delayed connection. The explicit declaration of delayed connection is necessary to ensure, without other knowledge of the processing objects internal behavior, that there is not a circular dependency between the data exchanged during a single timestep; this kind of dependency might otherwise occur when processing objects are connected in a loop.
Even after the removal of the delayed connections, deciding what is the best order of execution of the processing objects for an arbitrary network is a complex scheduling problem. We devised a relatively simple approach which does always provide a valid execution (or trigger) order under one single condition: once the delayed connections are removed, the resulting graph describing the simulation must be a DAG (Directed Acyclic Graph). This condition might appear to be too restrictive; however, it was always met in the modeling of the physical systems we have tackled so far: after all, it simply equivalent to forbid, in the model, loop of simultaneous dependencies between distinct objects. The precondition for a single processing object execution is that its inputs are available when its turn comes: note how this is guaranteed by the way the trigger order is predetermined.

\begin{figure}[t]
    \begin{center}
        \subfigure[\label{fig:triggerAssign1}]
        {\includegraphics[trim={0 4cm 0 0},width=0.49\columnwidth]{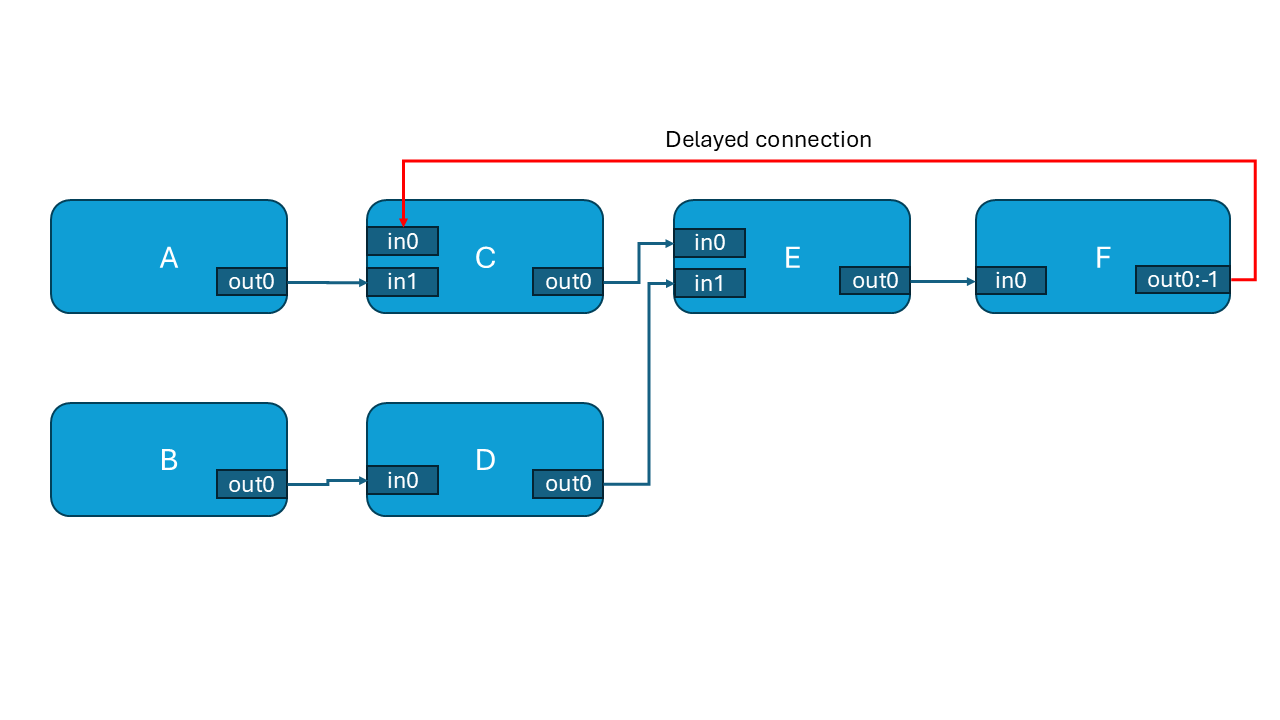}}
        \subfigure[\label{fig:triggerAssign2}]
        {\includegraphics[trim={0 6.8cm 0 0},width=0.49\columnwidth]{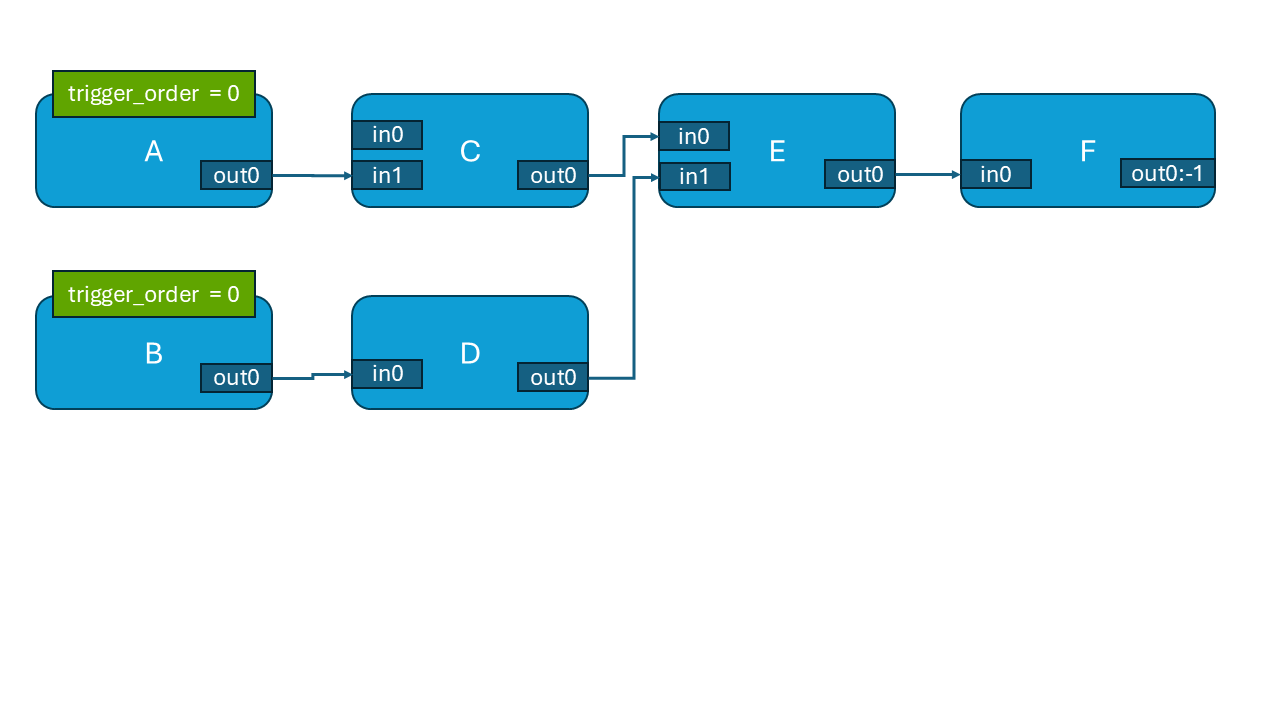}}
        \subfigure[\label{fig:triggerAssign3}]
        {\includegraphics[trim={0 6.8cm 0 0},width=0.49\columnwidth]{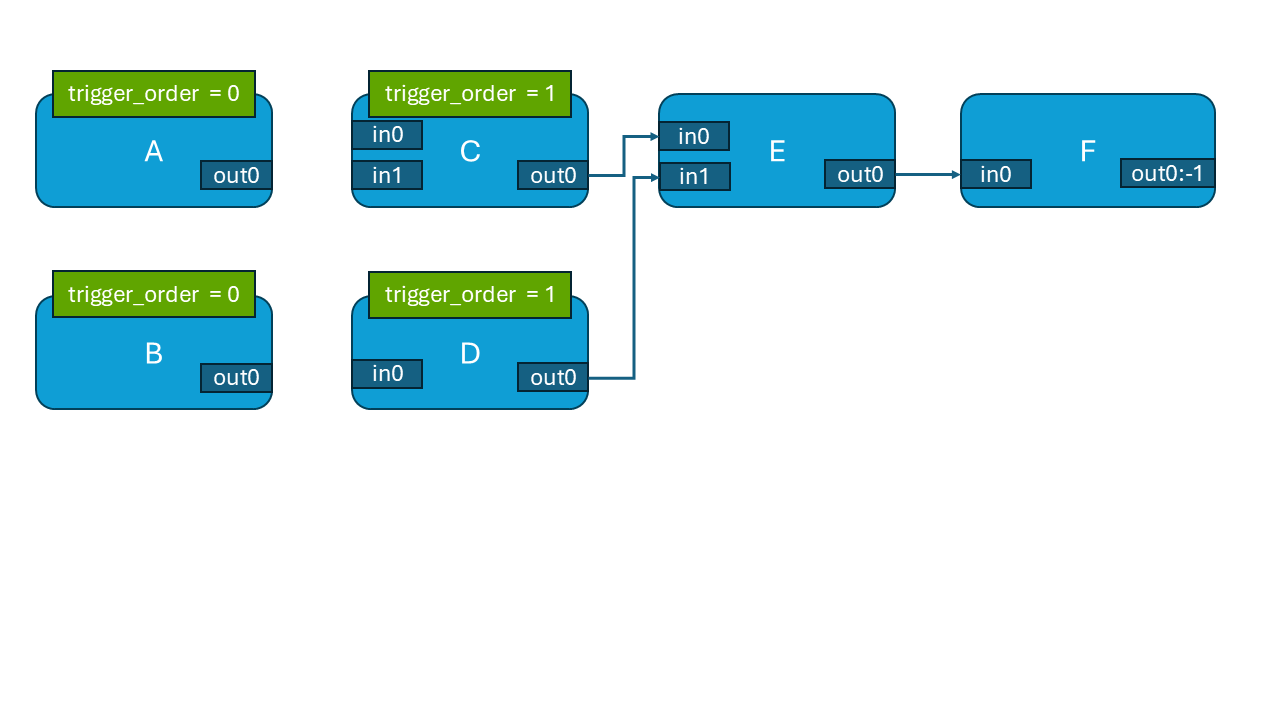}}
        \subfigure[\label{fig:triggerAssign4}]
        {\includegraphics[trim={0 6.8cm 0 0},width=0.49\columnwidth]{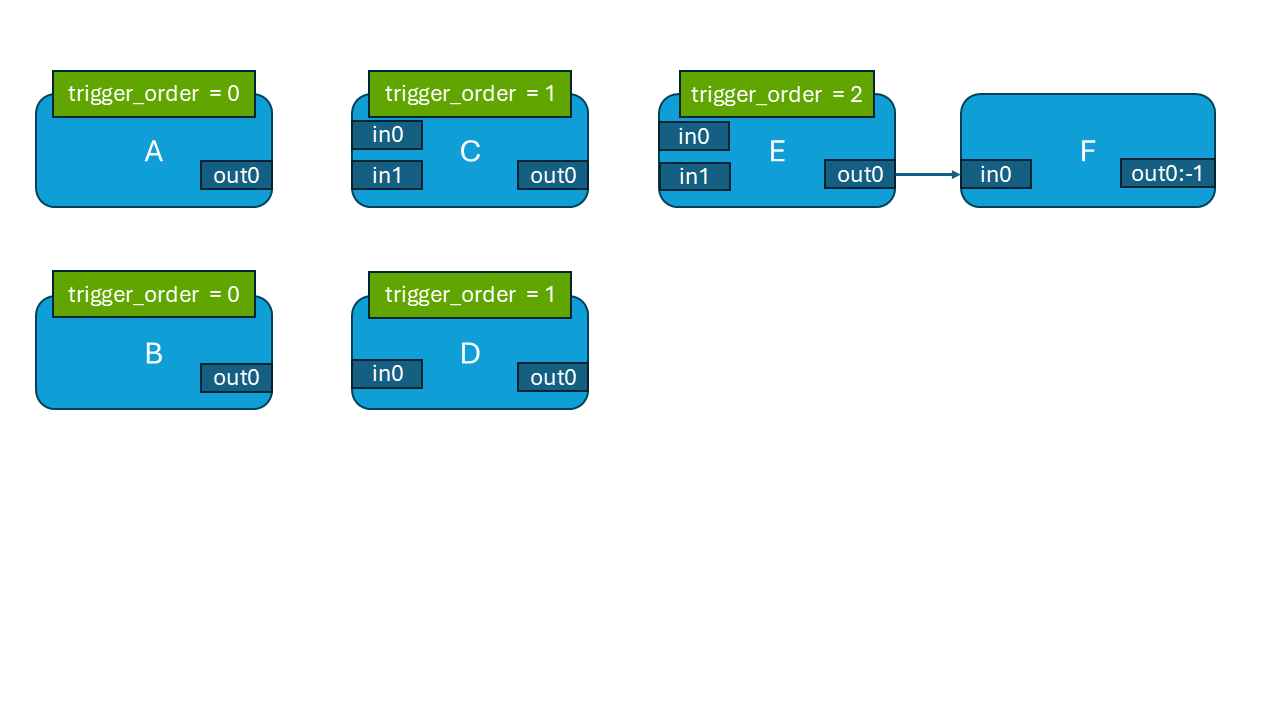}}
        \subfigure[\label{fig:triggerAssign5}]
        {\includegraphics[trim={0 6.8cm 0 0},width=0.49\columnwidth]{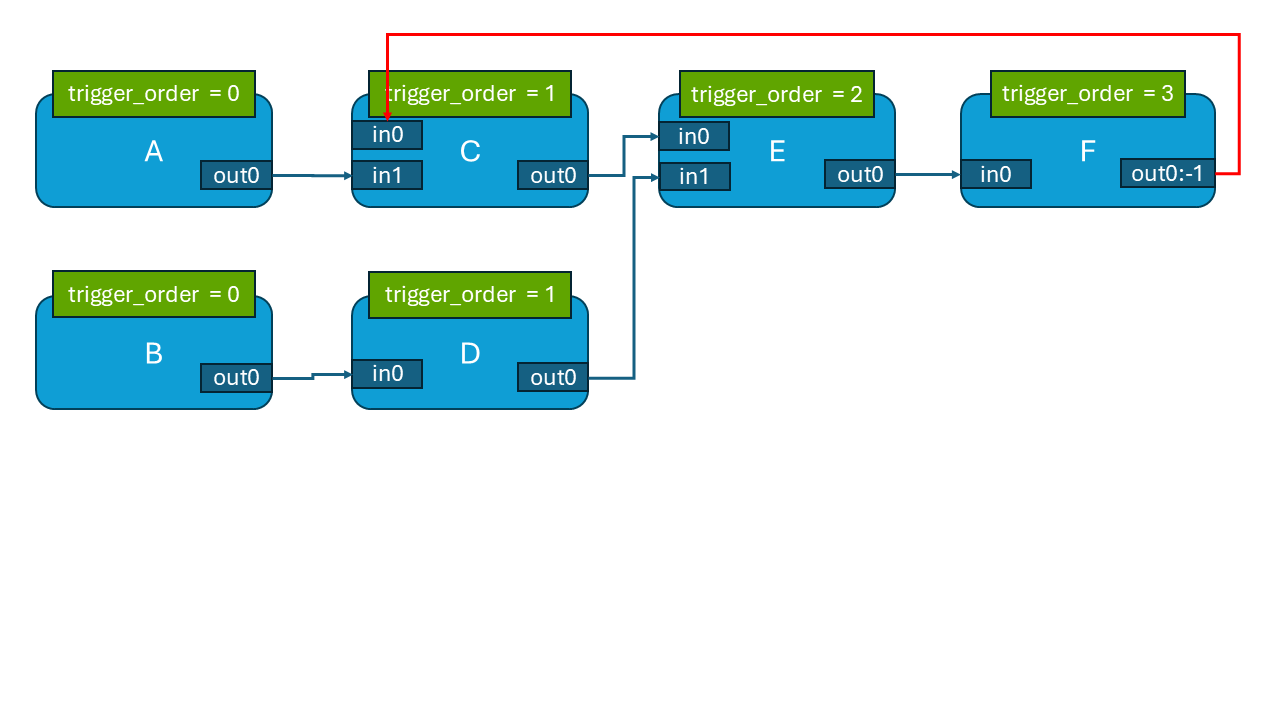}}
    \end{center}
    \caption{\label{fig:triggerAssign}The process of assigning a trigger order to each processing object in SPECULA.
    (a) Initial simulation graph. (b) Delayed connections are removed, leaves get the initial  trigger order (0). (c) Leaves are removed, trigger\_order is incremented, new leaves get assigned the current trigger\_order. (d) Leaves are removed, trigger\_order is incremented, new leaves get assigned the current trigger\_order. (e) Leaves are removed, trigger\_order is incremented, new leaves get assigned the current trigger\_order. All the connections are restored.}
\end{figure}

After verifying that the DAG condition is met, we assign to each node a trigger order, identifying in this way the classes of objects whose execution order is unconditioned. The trigger order assignment process is depicted in Fig. \ref{fig:triggerAssign}. We first assign 0 to the DAG leaves\footnote{To be consistent with the definitions from graph theory of root and leaf, consider the opposite direction of the graph arcs.}, which are the processing objects that do not have input (typically these are signal generators in our examples), and remove these nodes from the graph. This step is then repeated until no nodes are left in the graph. In the general case, the trigger order that is determined this way is only one of the possible valid ones, but it is always valid. This algorithm considers a single sequential executor (or process); in the case of multiple executors, the trigger order is still relevant for processing objects assigned to the same executor, while the cross-executor synchronization relies on the inter-process communication, as described in Sect.\protect\ref{sec:mpi}.

The entry point to use SPECULA is a command-line call to the program \texttt{specula}, which accepts a few options (see the documentation for details). This kind of entry point makes it possible to run in a multiprocess setup, for example using commands like \texttt{mpiexec} or \texttt{srun} from the \texttt{slurm}\cite{10.1007/10968987_3} scheduler.

\begin{figure}[t]
    \begin{center}
        \includegraphics[scale=0.5]{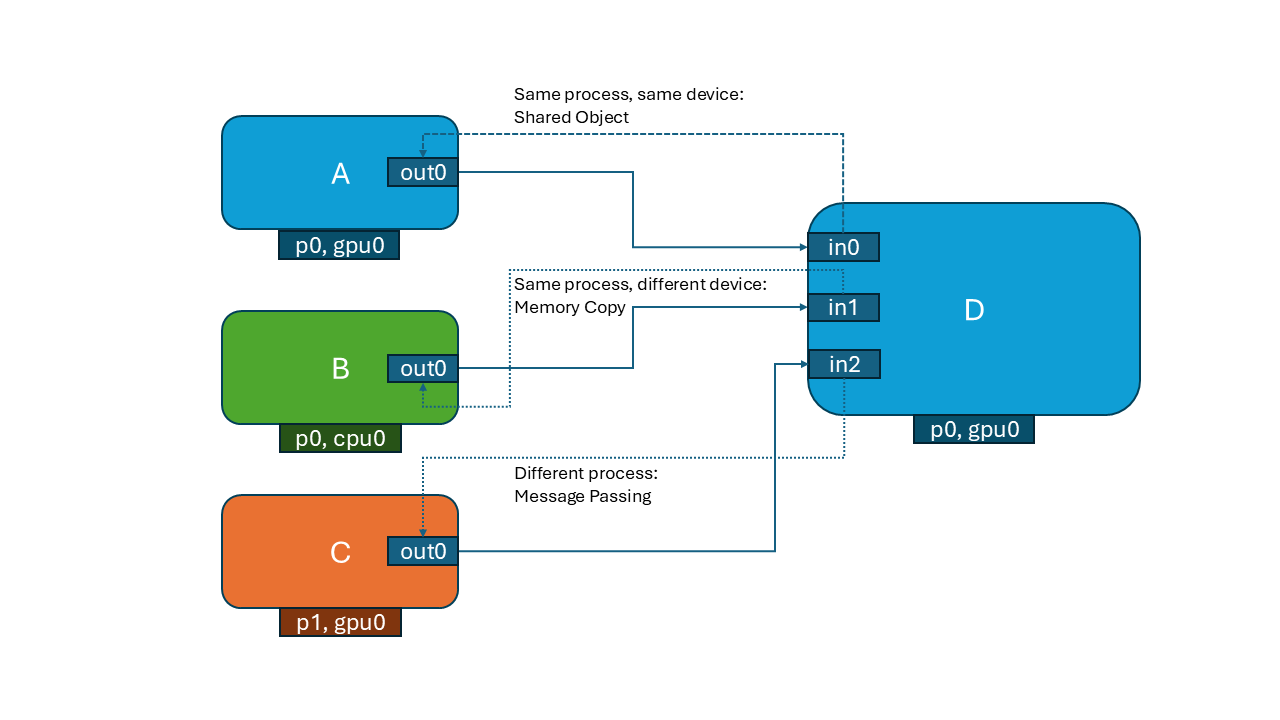}
        \caption{The inputs of D are handled as s simple reference to the corresponding output in the case of the output from A, with a memory copy in the case of the output from B, with a message passing mechanism in the case of the output from C.}
        \label{fig:connection}
    \end{center}
\end{figure}

\subsection{Data Classes}\label{sec:dataobjs}

Data objects represent the self-contained units of information exchanged between processing objects. All Data Objects inherit from a common \texttt{BaseDataObj} class, ensuring a consistent interface across the framework. The main functionalities they implement are memory allocation, basic manipulations, and data saving. Another key aspect of their design is temporal awareness: each object tracks its \texttt{generation\_time}, which is the time it was last modified. This allows us to skip unnecessary computations in a processing object when none of its inputs have been updated in the current time step.

The implementation is device-agnostic. A data object owned by a processing object as a local variable or as an output value is allocated to the same device and process of its owner. When a data object representing an output serves as input to a second processing object, the latter will have a reference to it. Since the two processing objects can reside on different devices or even in different computing nodes, the communication is handled automatically, depending on the case, relying on shared memory, memory copy or message passing mechanisms, as shown in Fig. \ref{fig:connection}. We call the latter case ``remote connection''.

Data persistence for all data objects is implemented in standardized \texttt{save} and \texttt{read} methods using the FITS format by default, ensuring that any simulation state, including all necessary metadata, can be archived, reloaded for analysis, or replayed for reproducibility. 
The framework provides a wide set of Data Classes which are typically needed in Adaptive Optics modeling, summarized in Tab.\ref{tab:dataObjects}. These classes can serve as base classes for future specializations, and their set is open to be extended. Please note that the Signal Processing category of Data Classes uses the Python Control Systems Library\cite{python-control2021}.

\subsection{Processing Objects}\label{sec:procobjs}

All processing objects inherit from a common \texttt{BaseProcessingObj} class. 
Their life cycle is strictly defined by initialization, input/output connection, setup, and repeated calls to the methods implementing their evolution during a time step. 
During initialization, objects are configured with default and/or specific parameters, pre-allocate on the target device (CPU/GPU) the needed working arrays using a unified interface (\texttt{self.xp}), and declare their typed inputs and outputs. 
Before simulation execution begins, a \texttt{setup} method finalizes each object's initialization, preparing for efficient GPU operation via CUDA\cite{NVIDIA2025} graph capture\footnote{Embedding parts of the CuPy code in  CUDA graphs improves performance by reducing the CPU/GPU communication overhead and allowing a higher occupancy rate of the GPU cores}, when applicable.
The computation performed at each time step is divided into these four phases.


\begin{itemize}
    \item \texttt{check\_ready}: the step demanded to perform a non-parallelizable (in the CUDA context) setup (like inputs gathering) and computation 
    \item \texttt{trigger}: the core computational part. The code might be written according to the rules that allow it to be captured into a CUDA graph, which will be executed into a CUDA stream. These rules include that no memory allocations and no host–device synchronization are allowed. In the \texttt{setup} method, an explicit call of the \texttt{build\_stream} method will perform the capture. 
    Developers who are not familiar with CuPy/CUDA graphs and streams or that are not willing to optimize the code to this point, can ignore these complications, while still implementing code that will run on GPU/CPU.
    \item \texttt{post\_trigger}: synchronization on GPU results and update of output objects.
    \item \texttt{send\_outputs}: send the outputs to processing objects that are allocated to different processes (only in the case of remote connections). 
\end{itemize}

This design aims at achieving computational efficiency while maintaining flexibility for both high-performance GPU and standard CPU execution. At the time of writing, we implemented more than 50 processing objects, most of which are listed in Table \ref{tab:procObjects}.

In addition to the code of the currently available processing objects, in the documentation we provide some details and guidelines to implement new ones.

\subsection{Management Objects}\label{sec:managementobjs}

These objects provide the backbone of the simulation handling, so we consider their extension or modification as more rare with respect to processing and data objects, for which instead we encourage the users of SPECULA to implement more classes in case of specific needs.

The \texttt{Simul} class serves as a container entity for the whole simulation, translating declarative yaml configurations into a set of organized and interconneted objects. It also manages the life cycle of the components of a simulation. It starts by parsing the description file(s) provided, combining the first description with the optional overrides to form the complete simulation configuration. Then it proceeds to instantiate the processing and data objects, resolving their interdependencies and references.
In the case of distributed simulation, a different  \texttt{Simul} object will be created for each process (identified by an integer index, \texttt{rank} in the MPI lexicon): the implementation of the same protocols and conventions allow the different \texttt{Simul} objects to know which parts of the simulations they have to manage and how the communication can be performed.
The class constructs the simulation's computational graph by establishing connections between object inputs and outputs, managing the setup of local Python references and remote MPI-based data transfers (in this case assigning a unique \texttt{tag} for the MPI communication). 
Connections management makes use of the \texttt{InputList} and \texttt{InputValue} classes, which can handle the case of shared memory communication (when connected objects live in the same computing node) or message passing (in the case of connected objects residing in two different computing processes). 

\begin{sidewaysfigure}[ph]
    \begin{center}
        \subfigure        
        {\includegraphics[scale=0.082]{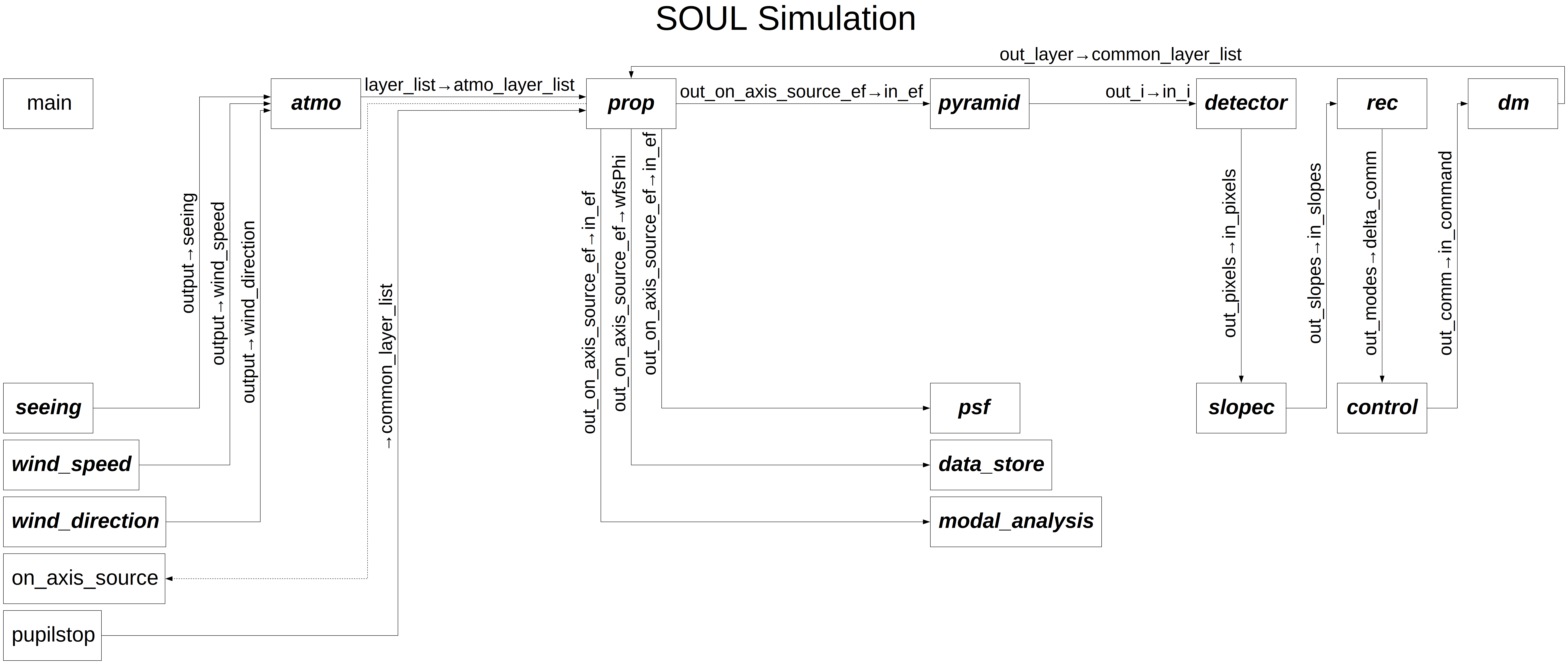}}
    \end{center}
    \caption{SOUL simulation diagrams. Diagram generated for the complete simulation, corresponding to the configuration file in Listing \ref{listTxt:soulConfig}. See also Fig. \ref{fig:SOULrepdiagram} for the corresponding replay diagram.\label{fig:SOULdiagram}}
\end{sidewaysfigure}

After the processing objects are built, connected, and assigned a trigger order, the \texttt{Simul} object initializes and controls the main simulation loop, optionally generating a visual diagram of the processing graph and supporting replay modes by transforming data-saving objects into data-loading ones.
For the generation of diagrams, we rely on the library \texttt{orthogram}\footnote{\href{github.com/yorgath/orthogram}{github.com/yorgath/orthogram}}.
A diagram can be generated highlighting the assignment of processing objects to various processes and GPUs (as in Fig.\ref{fig:MorfeoLeonardoColor}) or not (as in Fig.\ref{fig:SOULdiagram} or \ref{fig:SOULrepdiagram}).

The \texttt{LoopControl} class functions as the simulation scheduler and executor. Its primary role is to manage the repeated execution of the processing objects in accordance with the pre-defined trigger order. In the case of a single-process execution, this order is predetermined and static. The class initially orchestrates the \texttt{setup} phase for all objects, ensuring that proper memory allocation is performed and that MPI communication is setup for the inter-process data transfers. Its core \texttt{iter} method drives the main simulation cycle: for each time step, it sequentially processes groups of objects having the same trigger order, by calling on them the four \emph{time step phases} mentioned earlier ( \texttt{check\_ready}, \texttt{trigger}, \texttt{post\_trigger}, \texttt{send\_outputs} ).
This process ensures that objects are executed when their inputs are available and that the resulting data is promptly disseminated to dependent objects, including across MPI processes. The class also provides performance monitoring by reporting the simulation speed as execution frequency, and manages the finalization of all objects upon completion.

The \texttt{CalibManager} class is the centralized repository for managing persistent calibration and data files. It maintains an organized file hierarchy that maps specific data types to dedicated subdirectories. These include matrices, phase screens, pupil definitions, etc. The class provides a unified interface for constructing file paths and performing read/write operations using the FITS format, ensuring consistent data handling and storage.

The \texttt{FieldAnalyser} class can be used to extend the basic replay mechanism, see Sect.\ref{sec:store} for further details.

\section{Configuration files structure}\label{sec:conf}

The simulation is uniquely described by one (or more) yaml file(s), whose contents must be compatible with the SPECULA code. This means that the conventions that we will summarize in this section must be respected and that it is not possible to define new classes or to change the parameter list of a specific processing object just by writing this information into the yaml configuration file(s). In the SPECULA repository we provide several examples of different complexity, aiming to ease the comprehension of our conventions and to facilitate the writing of new configuration files.
Each section of the yaml file describes a processing or a data object which will be instantiated when the simulation is built. The order of the objects' definitions is not relevant: an object can be referred to before its description is provided. A detailed description of the syntax used is provided in \ref{sec:appendixB}.

All processing objects have two parameters, \texttt{target\_device\_idx} and \texttt{target\_rank}, in common. The first indicates the CPU (when set to -1) or GPU device index (when set to a non-negative integer) where the object will reside. The second contains the MPI rank of the process that will execute the object. As for most of object parameters, their specification in the yaml file can be omitted, thus relying on default values. We will talk more about these in Sect.\ref{sec:mpi}.

As mentioned earlier, a simulation description can be the result of the combination of multiple yaml files. This mechanism was introduced to allow incremental changes to the basic description of the system to be simulated, while avoiding the proliferation of configuration files. Typically, once we have specified the components of the simulation and their connections in a base file, we might want to run another simulation based on the same structure with some changes, for example if we have to perform a calibration, some components will be removed and other added. We might also want to keep the same structure but heavily manipulate the allocation of the objects to the various GPUs and/or processes. Also about this topic, more details can be found in the documentation; here we just recall that the SPECULA entry point of the simulation execution accepts one or more yaml files which are combined in a single structure. 

\section{Data Storage and Simulation Replay}\label{sec:store}

The final output of a simulation is typically some data that are stored on disk. In SPECULA this is handled explicitly by adding a specific processing object of class \emph{DataStore} to the simulation specification (the yaml description). The inputs to the \emph{DataStore} object will be saved as temporal series (the data plus the time axis), either in fits data format or as pickled Python objects.

\begin{sidewaysfigure}[ph]
    \begin{center}
        \subfigure
        {\includegraphics[scale=0.09]{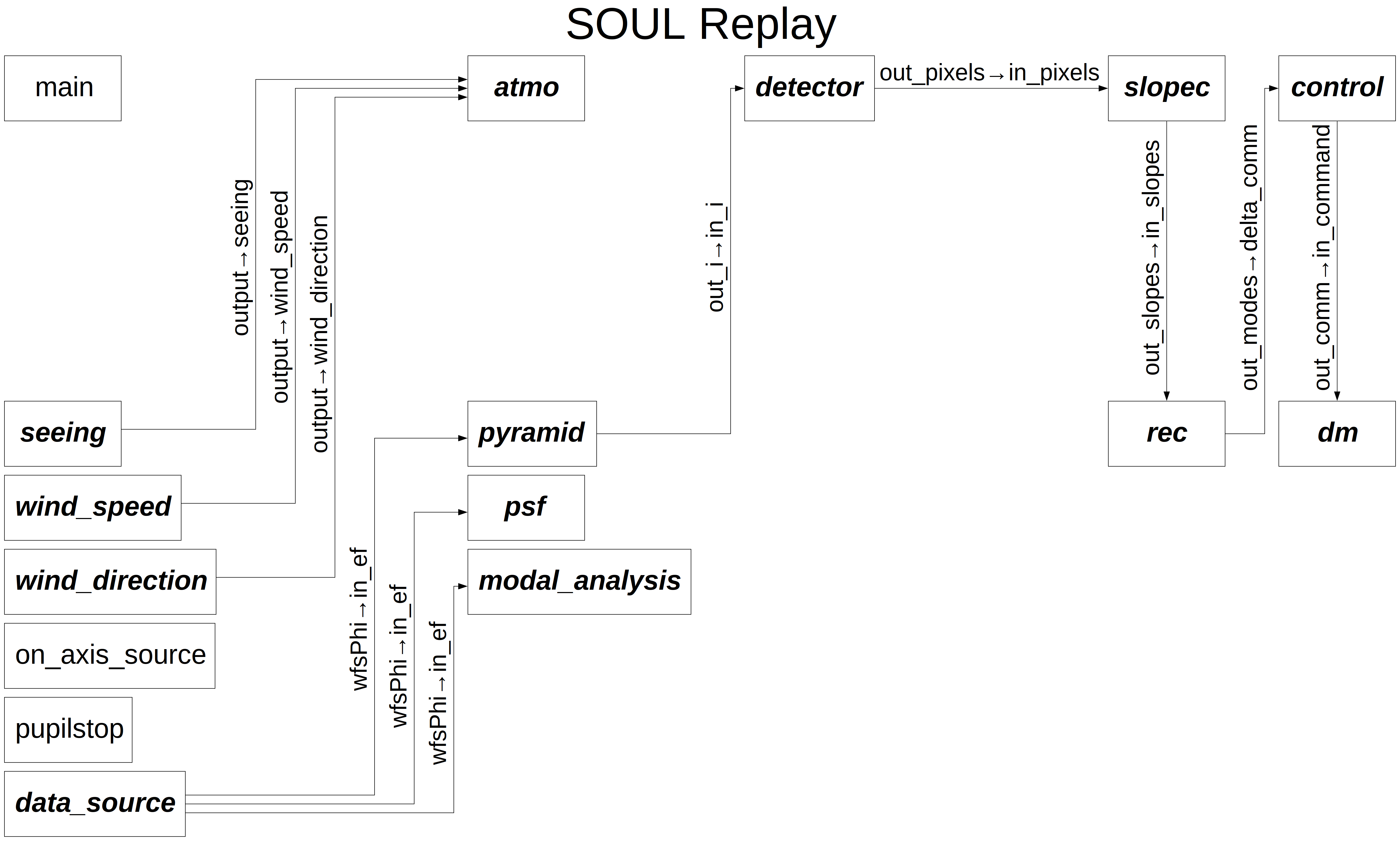}}
    \end{center}
    \caption{SOUL simulation diagrams. Diagram generated for the simulation replay.\label{fig:SOULrepdiagram}}
\end{sidewaysfigure}

The reader can look at a complete example in Listing \ref{listTxt:soulConfig}. The \emph{DataStore} will take care to create a new folder, under the provided \texttt{store\_dir}, as a ``tracking number'' folder named using the current system date and time in the format AAAAMMGG\_HHMMSS. At the end of the simulation run, the ``tracking number'' folder will contain one file for each of the inputs of the \emph{DataStore} object (something like \texttt{aFilename.fits}) and two additional files. The first is a parameter file named \texttt{params.yml} containing the full simulation structure: when multiple yaml files are provided as input, using the override mechanisms, this file is the result of their combination; this is meant to be able to reproduce the simulation in the future, making sure to have a track of the complete configuration used. The second is a file called \texttt{replay\_params.yml} and is created automatically to provide the functionality of replaying the simulation, by reloading the data from the tracking number instead of repeating the computations that were needed to generate such data. This is achieved by introducing an object of the class \emph{DataSource} in the replay file and removing the objects that have all their outputs saved in the tracking number. Note that replaying a simulation does not generate a new tracking number, as the \emph{DataStore} object is removed. The \texttt{replay\_params.yml} might be copied and manipulated to change this behavior, adding displays or other processing objects.
An example of the resulting diagram for replay of the SOUL case simulation is shown in Fig.\ref{fig:SOULrepdiagram}.

\emph{DataStore} and \emph{DataSource} objects can also be used in combination with \emph{FieldAnalyser}, to simplify the reuse of the output data of a simulation. This allows the user to compute PSFs, phase cubes, or perform modal analysis at a later time.
For example, this can be done at different wavelengths and in different lines of sight with respect to the ones originally used. Again, further details can be found in the documentation.

\section{Visualization Tools}\label{sec:visual}

There are two different mechanisms to visualize the different data objects exchanged by the processing objects during the simulation evolution, Display Objects and Display Server. These two mechanisms can be used concurrently.

\begin{figure}
    \begin{center}
        \subfigure[{}]
        {\includegraphics[width=0.34\columnwidth]{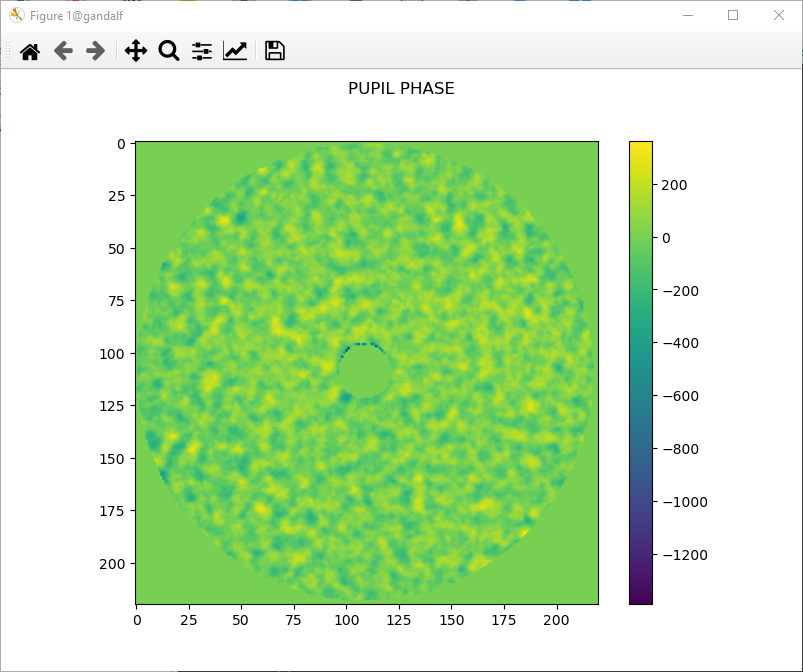}}
        \subfigure[{}]
        {\includegraphics[width=0.34\columnwidth]{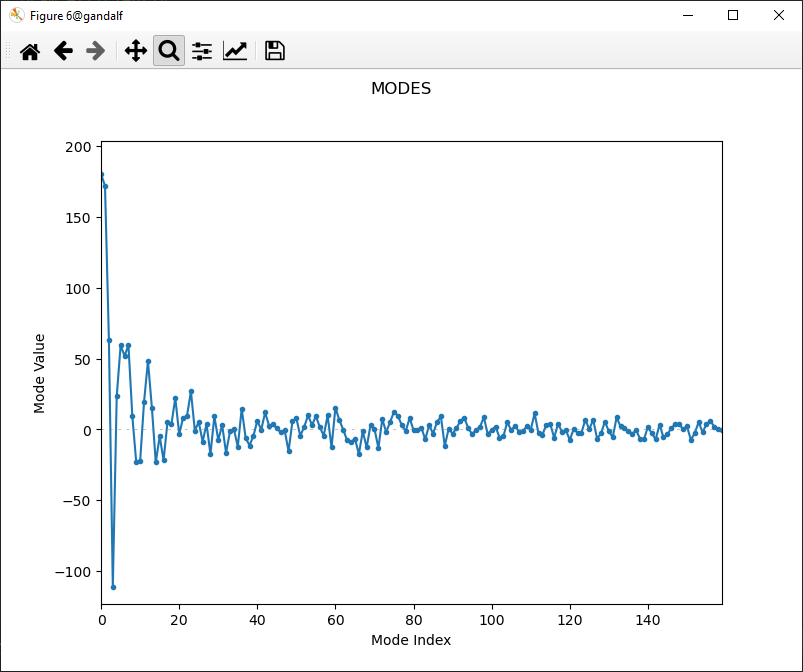}}
        \subfigure[{}]
        {\includegraphics[width=0.32\columnwidth]{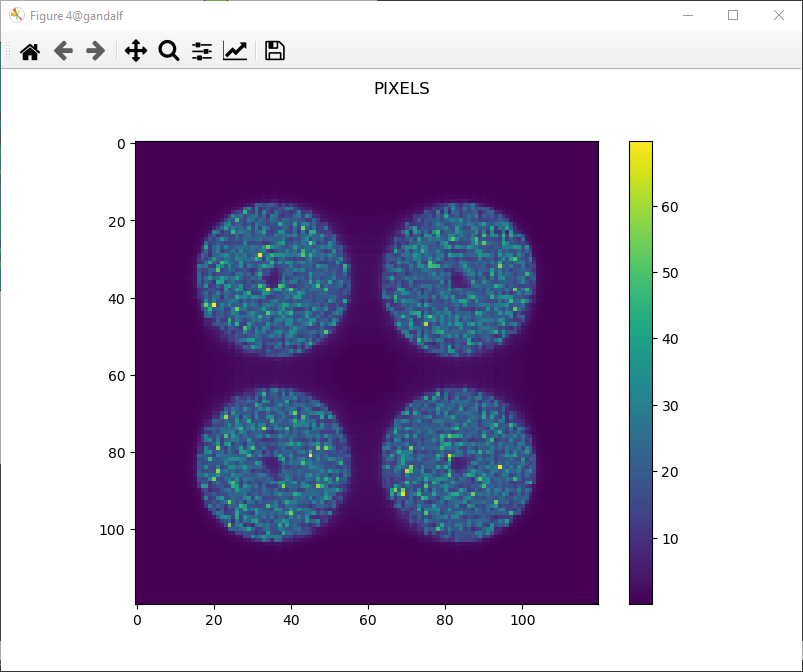}}
        \subfigure[{}]
        {\includegraphics[width=0.32\columnwidth]{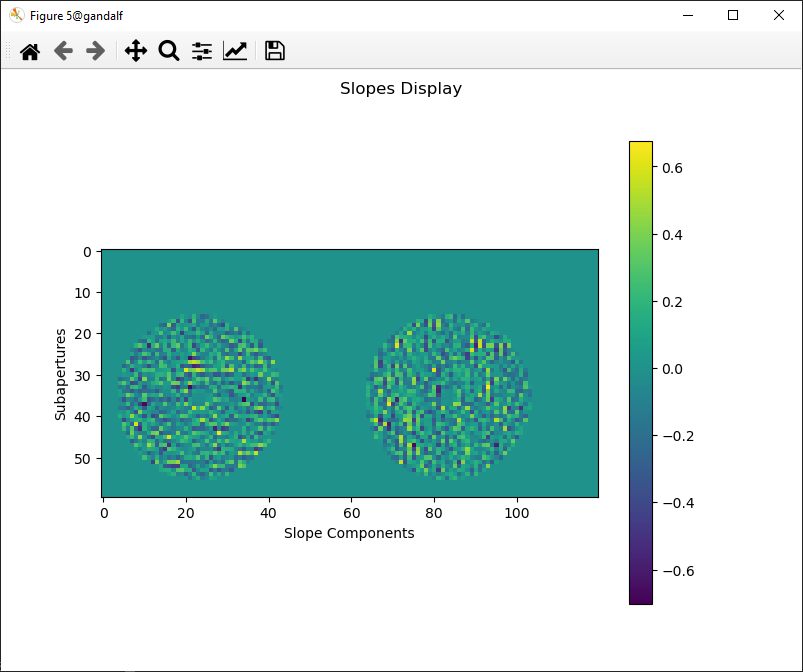}}
        \subfigure[{}]
        {\includegraphics[width=0.32\columnwidth]{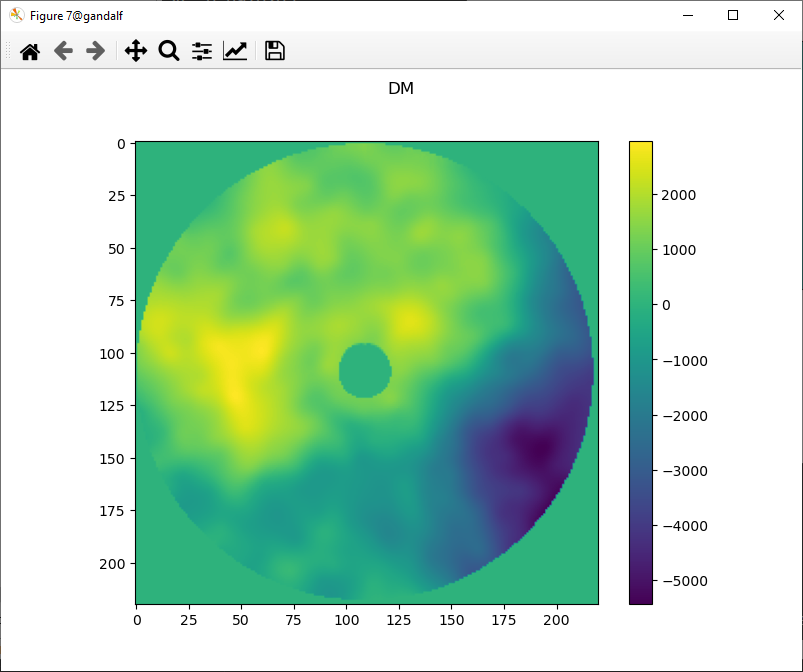}}
        \subfigure[{}]
        {\includegraphics[width=0.34\columnwidth]{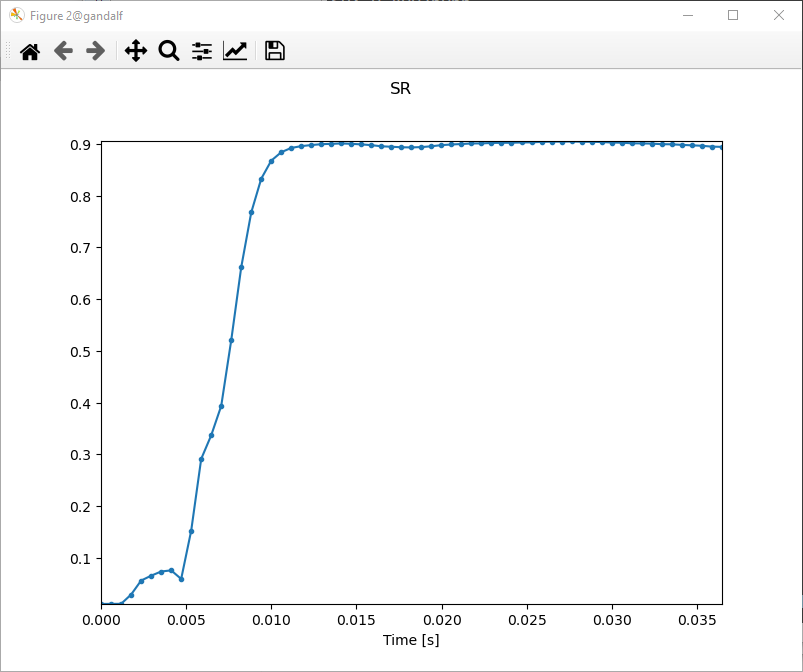}}
        \subfigure[{}]
        {\includegraphics[width=0.34\columnwidth]{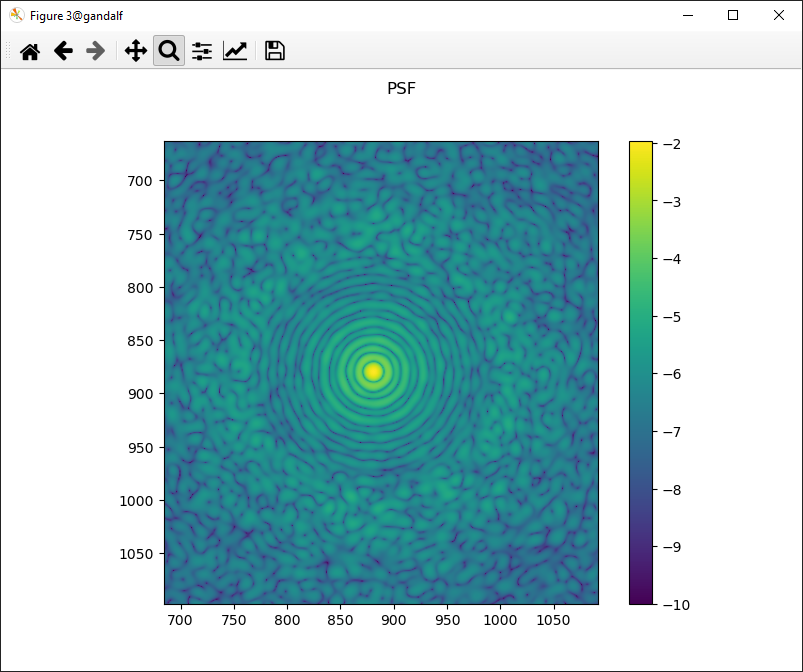}}
    \end{center}
    \caption{A gallery of display objects output.
    (a) PhaseDisplay, at pupil. (b) ModesDisplay. (c) PixelsDisplay. (d) SlopecDisplay. (e) PhaseDisplay, at Deformable Mirror output. (f) PlotDisplay. (g) PsfDisplay.}
    \label{fig:displayObjects}
\end{figure}

\subsection{Display Objects}
We implemented a set of processing objects to provide the basic functionality of live display during simulation. The visualization is made relying on \texttt{matplotlib}\cite{Hunter:2007} library, and the general graphics handling is implemented in the base class \texttt{BaseDisplay}. This class was specialized into display classes for a scalar or vector time series, for 2D images, and for most of the data objects: Pixels, Slopes, Electric Field. 
The display objects have inputs to define the data they are going to display and can do some data preparation before the display, but differently from other processing objects, they have no outputs, as their output is considered to be the visualization itself.
Some examples of the displays are shown in Fig.{\ref{fig:displayObjects}}.

\begin{figure}[t]
    \centering
        \includegraphics[scale=1.5]{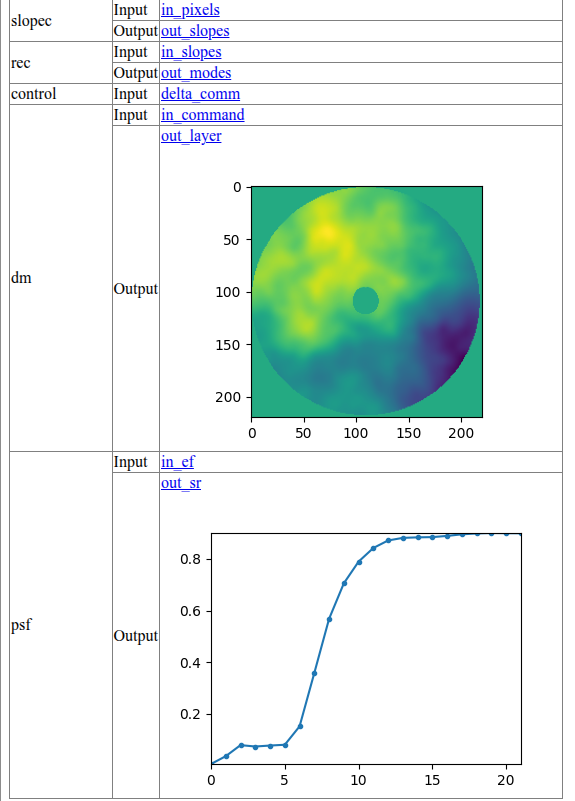}
        \caption{Example a web page generated by the display server, showing the output of two processing objects after clicking on the ``out\_layer'' and ``out\_sr'' links.}\label{fig:webDisplay}
\end{figure}

\subsection{Display Server}

SPECULA can set up a Flask web server to transmit the displays over the network. The display server can be added as a processing object to a simulation, and it spawns the web server via the multiprocessing module. Web clients connecting to it will be provided with a Javascript page using the SocketIO library (\href{socket.io}{socket.io}). The Javascript code is able to parse a configuration dictionary derived from the simulation yaml file, and allows the user to interactively request a subset of the data generated by the simulation in real-time, by opening or closing a dedicated <div> element for each input or output of each processing object. The requested data are rendered internally by the display server at each simulation iteration, as images or plots, and sent to the web client as PNG images.
The resulting display is non-interactive and has some limitations, like a fixed range for most elements, but is useful as a simulation quick look or sanity check. An example screenshot of the web page is shown in Fig. \ref{fig:webDisplay}.

\section{Multiprocessing and HPC enviroment}\label{sec:mpi}

Because SPECULA supports MPI, a simulation can be distributed across several computing nodes. The implementation of this support is based on the \texttt{mpi4py}\cite{Dal05} library, which wraps the Message Passing Interface (MPI)\cite{Gro99a}, a portable message passing standard designed to work on parallel computing architectures. We already mentioned how a processing object can be assigned to a GPU of a specific index, among the available ones, using the parameter \texttt{target\_device\_idx}. The allocation to a specific process is made in a similar way by assigning an integer value to the parameter \texttt{target\_rank}.
Note how the user cannot directly assign a processing object to a computing node, however, this can be achieved in an indirect way, since the process scheduler available in the HPC environment is usually configured to assign, following a consistent process indexing, a fixed number of processes to each node. For example, if this number is 4, and we are going to use 3 computing nodes, we know that process ranks in 0..3 will be assigned to one node, process ranks in 4..7 will be assigned to another node and so on. GPU indexing stays relative to the node on which the processing object resides, so if each of our computing nodes has 4 GPUs and we want each process to use a different GPU, we would use the following pairs of \texttt{target\_device\_idx} and \texttt{target\_rank}: (0,0), (1,1), (2,2), (3,3), (0,4), (1,5), (2,6), (3,7), (0,8), (1,9), ... (3,11).

\begin{sidewaysfigure}[p]
    \begin{center}
        \subfigure        
        {\includegraphics[ width=0.9\columnwidth]{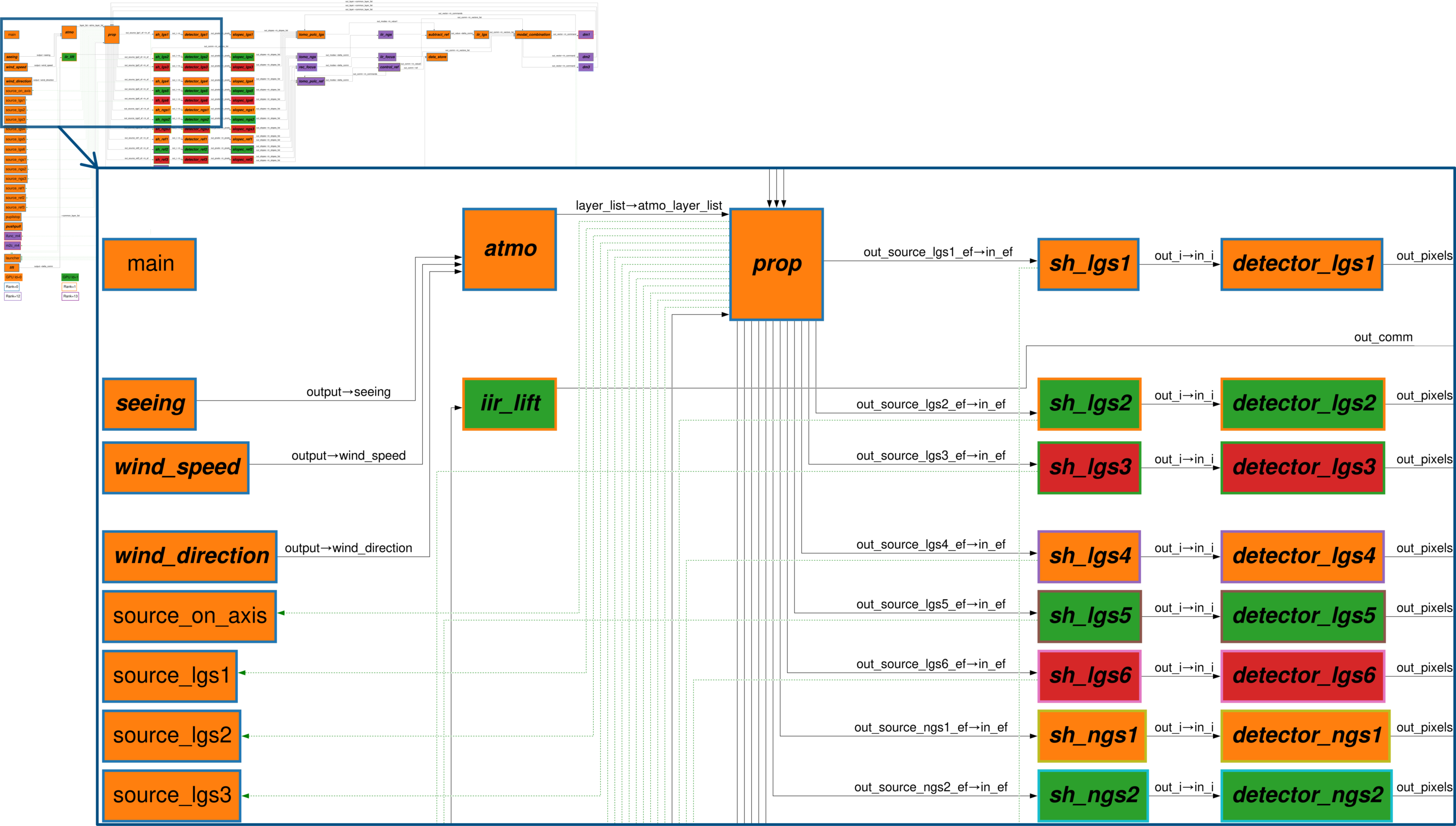}}
    \end{center}
    \caption{A crop of the huge diagram for the Full MORFEO simulation, allocated on 4 HPC nodes. }\label{fig:MorfeoLeonardoColor}
\end{sidewaysfigure}

Process allocation and GPU indexing are also supported in diagram production, by color coding of the border and filling of the blocks representing the objects. An example is shown in Fig.\ref{fig:MorfeoLeonardoColor}.

As already mentioned in Sect.\ref{sec:basicconcepts}, when two objects are allocated to two different processes, their connection is implemented using the MPI: more specifically, we rely on a non-blocking send by the output owner and on a blocking receive on the input owner side: this means that the processes hosting the latter object will be suspended by the operating system until the data are received. This is the only synchronization mechanism between various processes because the execution order based on the trigger order is respected only within the processing objects assigned to the same process. As we tested, it is then possible to allocate each processing object to a different process (even on a single CPU/GPU machine). In this case, from a scheduling point of view, we simply rely on the operating system and the general partial order relationship between objects (shown in Fig.\ref{fig:partailorder}): as a consequence, depending on the relative computation time required by the different objects, all the possible, valid, execution orders might be explored. For relatively simple systems, such as SOUL or SCAO systems in general, being able to distribute the processing objects across several processes and GPUs does not provide an advantage, given the intrinsically sequential nature of the computation to be performed at each time step. For other more complex systems as MORFEO\cite{2024SPIE13097E..22C} or other Multi-Conjugated AO (MCAO) systems, just looking at the simulation diagram makes it apparent which groups of processing object could be run in parallel by multiple executors, with a consequent gain in performances.

\begin{table}[t]
    \centering
    \begin{tabular}{|l|c|c|c|c|c|} 
    \hline 
    Simulation & HPC nodes & CPUs & GPUs & Speed[Hz] & Time step [ms] \\
    \hline 
    Ref: SOUL in PASSATA & 1 & 1 & 1 &  98.73 &  10.1 \\
    \hline 
    SOUL & 1 & 1 & 0 &  1.04 &  961.0 \\
    \hline
    SOUL & 1 & 1 & 1 & 162.2 & 6.2 \\
    \hline
    SCAO Shack-Hartmann & 1 &  1 & 2 & 169.5 & 5.9 \\
    \hline
    ANDES & 1 & 1 & 1 &  19.6 & 50.9 \\
    \hline
    ERIS & 1 & 1 & 2 & 77.02 &  13.0 \\
    \hline
    MORFEO & 1 & 1 & 4 & 7.81 & 128.1 \\
    \hline
    MORFEO & 2 & 2 & 8 & 12.76 & 78.36 \\
    \hline
    MORFEO & 4 & 4 & 16 & 13.19 & 75.82 \\
    \hline
    MORFEO + \texttt{DistributedSH} & 7 & 7 & 28 & 23.85 &  41.9 \\
    \hline 
    \end{tabular}
    \newline
    \caption{Performances in different use cases.}
    \label{tab:tablePerf}
\end{table}

Due to the fact that the MPI library is agnostic about the underlying architecture, we were able to develop, test, and debug the multiprocessing support on a simple workstation. We then tested SPECULA in HPC environment on the Italian super-computer LEONARDO \cite{2023arXiv230716885T}. To make use of the available computational power, we run the full simulation for MORFEO\cite{2024SPIE13097E..22C}, the more complex and computationally demanding Adaptive Optics system we are aware of. 
The batch scheduler that allows the access to Leonardo's nodes is the tool \texttt{slurm}\cite{10.1007/10968987_3}: in Listing \ref{listTxt:slurm} we show its configuration to run SPECULA over 4 HPC nodes. Each computing node in Leonardo is equipped with 4 Nvidia A100 GPUs (Custom Ampere A100, 64GB HBM2).

In MORFEO we have 12 WFSs\cite{2023aoel.confE.129B} and each of the corresponding pipelines (WFS, detector and slope computer) can be executed in parallel on a different GPU. The computational needs of the different WFS varies, being Shack-Hartmann WFSs demands depending on their geometry. 

\begin{table}[t]
    \centering
    \begin{tabular}{|l|c|c|c|c|c|c|} 
    \hline 
    \multirow{3}{*}{Simulation}  & no. pixels & no. & WFS  & WFS FoV
& WFSs FFT & Total \\
    {} & on pupil  & sub- & FoV  & oversize & oversamp. & WFSs FFT \\
    {} & diam. & apertures  & [arcsec] & factor & factor & workload \\    
    \hline 
    \hline 
    SOUL & 220 & 40 & 2.1 &  1 & 3 & 24x660x660 \\    
    \hline
    SCAO SH & 160 & 20 & 2.4 & 2 & 2 & 314x56x56 \\
    \hline
    ANDES & 400 & 100 & 2.1 &  1 & 4 & 24x1600x1600 \\
    \hline
    ERIS & 160 & 140 & 5.1 & 2 & 2 & 1256x68x68 \\
    \hline
    MORFEO & 480 & 68 & 16.1 & 1.5 & 2 & 6x3630x240x240 \\
    \hline
    \end{tabular}
    \newline
    \caption{Simulation parameters describing the computational workload.}
    \label{tab:tableSizes}
\end{table}

We show in Tab.\ref{tab:tablePerf} the performances obtained for several AO systems and for different allocations of the relative simulations on the available computational resources, reporting the average time needed to perform all computation in a time step and the equivalent speed, as a frequency of execution. From the two allocations of SOUL is apparent how SPECULA really benefits from GPU acceleration: as a general rule, all of the computationally demanding processing objects should be allocated to the available GPUs. Other simulations performances largely depend on the size in pixels of the propagated electric field, which for sampling needs is higher in ELT than in VLT era system. ERIS can naturally be allocated to two GPUs, as it is composed by one Natrural Guide Star (NGS) and one Laser Guide Star (LGS) WFS. For MORFEO we experimented with various allocations, we report about the more significant ones. Going from using a single HPC node to two HPC nodes, we observe fairly good scaling, which is, however, not proportional: we have 12 WFSs of different computational requirements, besides the expected communication overhead. Going from 2 to 4 HPC nodes we are hitting a wall of how much the architecture of this specific case is parallelizable. To achieve the performances shown in the last row of the table, we had to go beyond the paradigms described so far regarding the processing objects allocation, adding the possibility of distributing the computations of a single processing object across the GPUs of a node. This extension was implemented in the \texttt{DistributedSH} class. In this case, each of 6 HPC nodes was entirely dedicated to each of the 6 LGS pipelines, while the rest of the simulation was allocated to the 4 GPUs on the 7th node. More details on the simulation sizes, from which it is possible to evaluate the computational load of the various cases, are provided in Tab.\ref{tab:tableSizes}. Note the columns about oversize and over-smapling factors which refer to padding and up-sampling operations and affect the final FFT shapes and computational load.

\section{Hardware in the loop}\label{sec:hwloop}

\begin{figure}[t]

    \begin{center}
        \subfigure[\label{fig:SPECULA_lab}]
        {\includegraphics[trim={0 15cm 0 15cm},width=0.9\columnwidth]{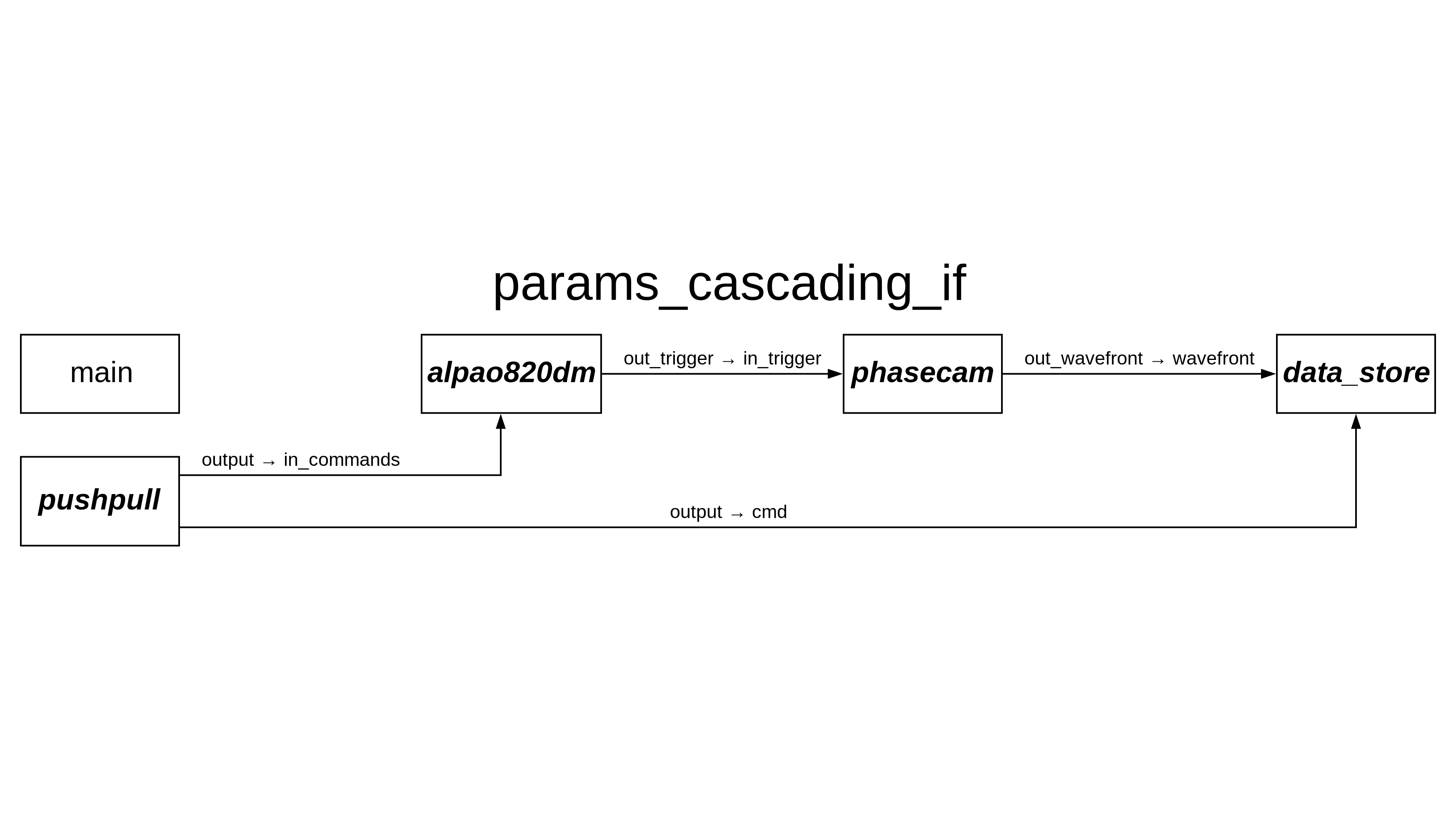}}
        \subfigure[\label{fig:SPECULA_lab_picture}]
        {\includegraphics[width=0.48\columnwidth]{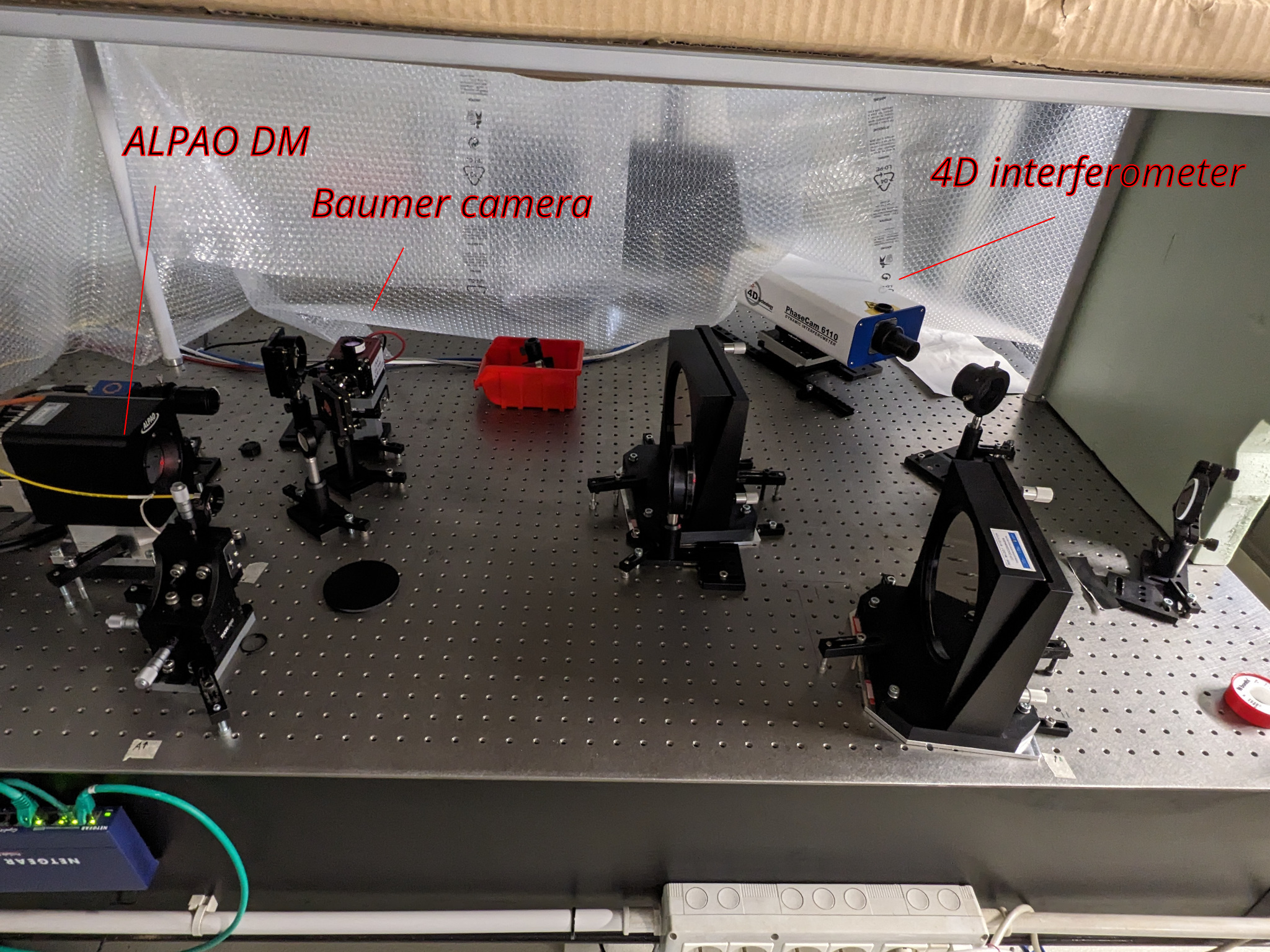}}
        \subfigure[\label{fig:SPECULA_IF_meas}]
        {\includegraphics[width=0.48\columnwidth]{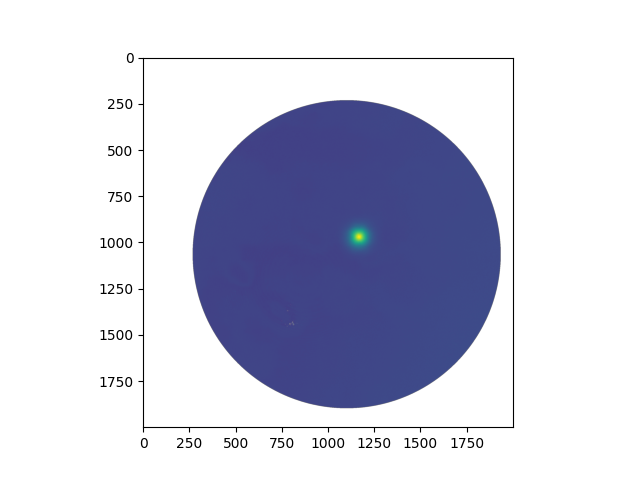}}
    \end{center}
    \caption{\label{fig:SPECULA_bench} Using SPECULA for Influence functions measurement of a physical DM.
    (a) SPECULA Diagram for the lab setup. (b) Optical bench picture. (c) Display of the measurement of a single influence function.}

\end{figure}

One of the key objectives of SPECULA is to support laboratory activity.
In this case, SPECULA is interfaced with the hardware by building specialized \emph{processing} objects that wrap the interface details and can communicate with the rest of the simulation, exchanging standard SPECULA data objects.

In our implementation, hardware is accessed using the PLICO\cite{2023aoel.confE..41S} framework, and thus any of the supported HW models available in PLICO are automatically available in SPECULA as well, with a very simple configuration, typically just the hostname and port of the relevant PLICO server. The SPECULA processing objects will connect to the PLICO server, issue a method call to get the data to or from the HW device (e.g. read from a camera or set a DM shape) and adapt the data format. At the moment, four different processing objects are available:
\begin{itemize}
    \item PlicoDM: can command a DM to an arbitrary shape, optionally applying a modal projection.
    \item PlicoInterferometer: reads wavefronts from a supported interferometer.
    \item PlicoMotor: sets and/or get motor position.
    \item PysilicoCamera: get a frame from a detector.
\end{itemize}

Each of the objects performs a single operation (e.g. read a camera frame, or set a motor position) for each simulation step. In this first version of the hardware support, we put the implementation of these objects in a separate repository, available at \href{github.com/ArcetriAdaptiveOptics/SpecuLab}{github.com/ArcetriAdaptiveOptics/SpecuLab} (we might add these objects to SPECULA and PLICO as an optional dependency).

One full example of hardware-in-the-loop with SPECULA is shown in Fig. \ref{fig:SPECULA_bench}, with the goal to calibrate of actuators' influence functions of an ALPAO mirror. The picture in \ref{fig:SPECULA_lab_picture} shows an optical bench with: an ALPAO DM, a Baumer camera, and a 4D PhaseCam interferometer. A simple SPECULA simulation (corresponding diagram shown in Fig. \ref{fig:SPECULA_lab}) drives the DM with a push-pull generator, reads wavefront images from the interferometer and saves both DM commands and wavefronts to disk, realizing a basic Influence Function measurement, of which an example result is shown in Fig. \ref{fig:SPECULA_IF_meas}. More complex use cases might use the same HW devices to apply turbulence, to reconstruct the modes and to close different kinds of hybrid simulation loops.

\section{Simulation results validation}\label{sec:compare}

\begin{figure}[h]
    \begin{center}
        \subfigure[\label{fig:comparison_SOUL_SRH}]
        {\includegraphics[width=0.48\columnwidth]{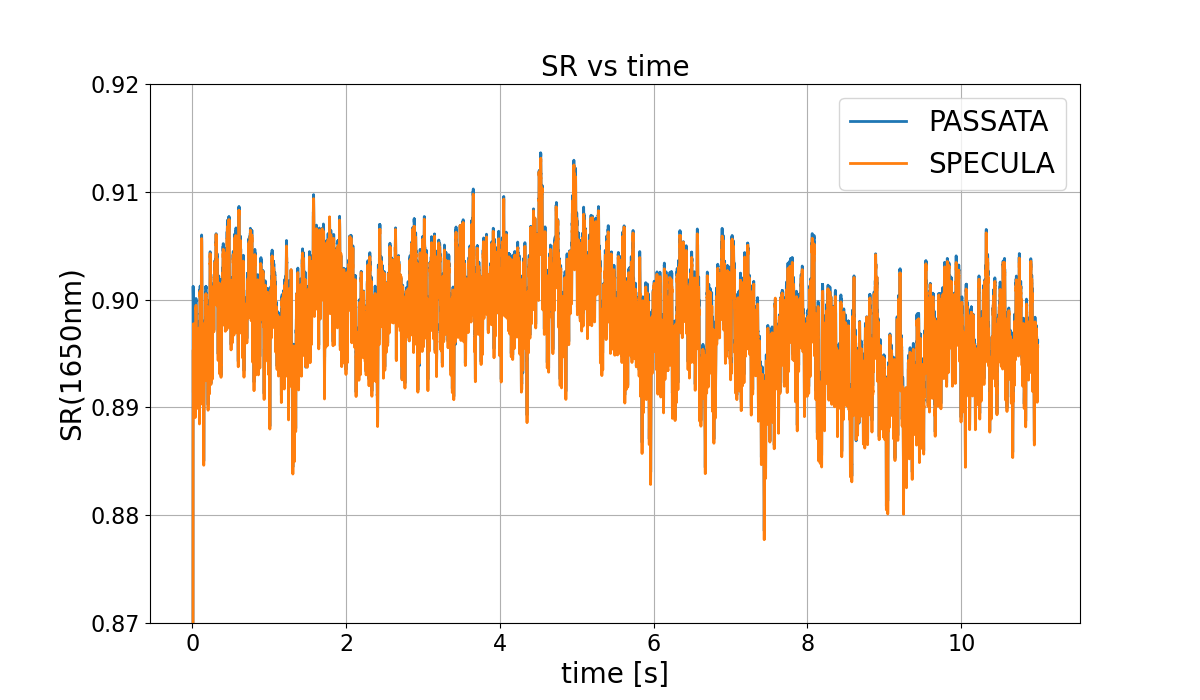}}
        \subfigure[\label{fig:comparison_SOUL_modal_plot}]
        {\includegraphics[width=0.48\columnwidth]{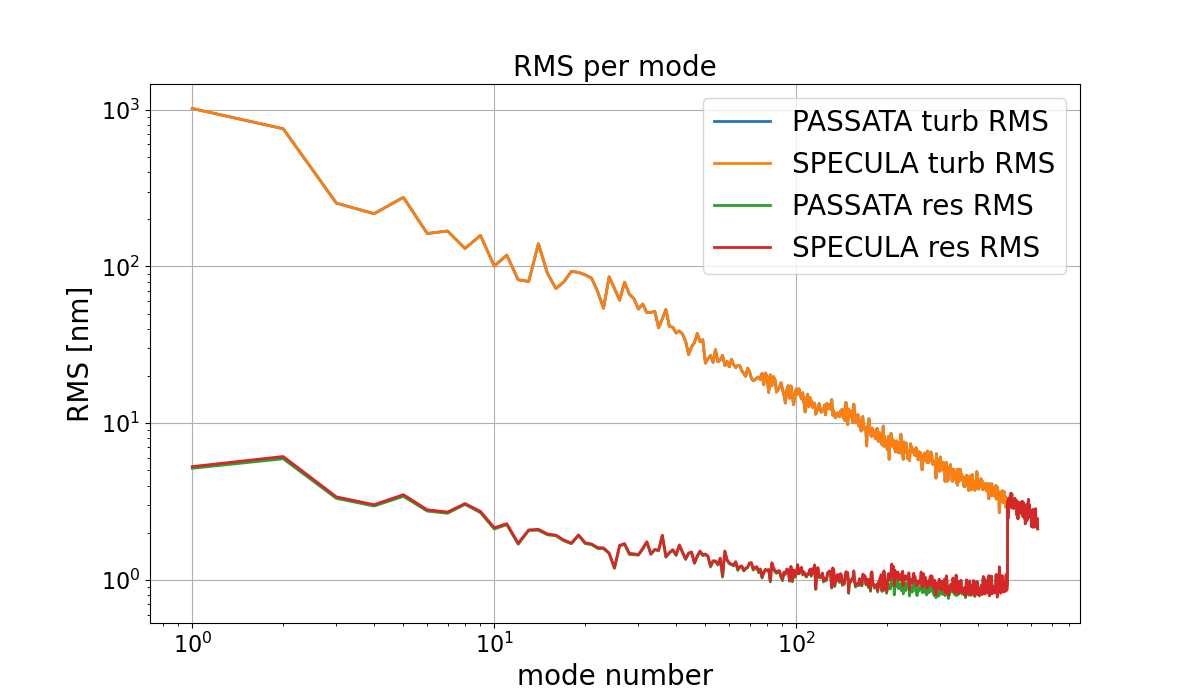}}
    \end{center}
    \caption{\label{fig:comparison_SOUL}Comparison of PASSATA and SPECULA performing a SOUL simulation of 11 s (guide star magnitude is 8 and seeing on the line of sight is 0.8 arcsec).
    (a) H band (1650 nm) Strehl Ratio. (b) Turbulence and residual modes RMS.}
\end{figure}

We followed a dual approach to validate SPECULA: internal software verification and external scientific validation against an established simulator.

Internal verification ensures the correctness and robustness of the code through a comprehensive test suite.
This includes unit tests to verify the individual functionalities of the software, integration tests to ensure correct interaction between modules, and validation tests that compare the output of specific components against reference data pre-computed with PASSATA.
The suite comprises more than 500 tests, distributed in approximately 100 files.
These tests are automatically executed whenever the code is modified, creating a continuous integration framework that prevents regressions and guarantees reliability.

To validate the results scientifically, we systematically compared SPECULA's results with analogous data obtained using PASSATA.
We consider PASSATA to be a reliable benchmark because it is a well-established simulator whose results have been verified in a variety of contexts, ranging from instrument design to on-sky commissioning.
As an additional argument for the validity of the simulation results of PASSATA, we recall how its outputs have been cross-checked against those of other simulators\cite{2019JATIS...5d9001A,2022SPIE12185E..3LA,2024SPIE13097E..2YN} and real-world astronomical data\cite{2010SPIE.7736E..09E,2011SPIE.8149E..02E,2021arXiv210107091P,2022SPIE12185E..08R}.

The validation of the results from SPECULA against the results from PASSATA was  performed for the entire set of AO systems listed in Table \ref{tab:tablePerf}. As a representative example, we present the results of a SOUL simulation.
Fig.\ref{fig:comparison_SOUL} shows the results of an 11-second simulation in the case of SOUL\cite{2023aoel.confE..80P} run on both SPECULA and PASSATA under identical conditions.
The agreement between the two simulators is excellent, with negligible differences attributable to numerical precision. 
Similar levels of agreement were found for all other systems that were tested.



\section{Conclusions and Future Work}\label{sec:conclusion}

After more than a year of development, we believe that SPECULA has reached the necessary level of maturity to be introduced to the Adaptive Optics community. This belief is based on the results we obtained in terms of performances, ease of extension and more importantly on the correctness of the simulations results, which we verified against the ones obtained in PASSATA for a wide set of systems. Having defined the core structure of the framework in a modular way, it is natural to think about a wide range of extensions: we are planning to extend the support for hardware in the loop operation and to add tools to make the simulation definition phase more user-friendly: a simple design GUI could help the user produce the yaml definition file for the simulation. The visualization tools can be extended, since different kind of GUIs can be easily implemented for the real-time feedback and analysis of the results. The chapter of hybrid simulation can also be expanded in many directions, adding support for a wider range of devices, and setting up many different use cases that arise in laboratory activities.
We encourage adoption and contribution from the community and we plan to keep the framework up to date, making it able to simulate the state-of-the-art Adaptive Optics systems, adding the needed processing blocks to achieve this goal.

\subsection* {Acknowledgments}
The authors wish to thank all the members of the Adaptive Optics group at the Arcetri Astrophysical Observatory for their support and useful discussion and suggestions. Special thanks to Marco Bonaglia and Edoardo Bellone De Grecis for their support in the laboratory tests. We acknowledge ISCRA for awarding this project access to the LEONARDO supercomputer, owned by the EuroHPC Joint Undertaking, hosted by CINECA (Italy). 

\section* {Disclosures}

The authors declare there are no financial interests, commercial affiliations, or other potential conflicts of interest that have influenced the objectivity of this research or the writing of this paper.

\subsection* {Materials and Methods}
We acknowledge the assistance of Overleaf AI Assist powered by Writefull for language refinement and formatting in the preparation of this manuscript.
For the laboratory setup we used an ALPAO 820 Deformable mirror, a 4D PhaseCam 6110 interferometer, Baumer VCXG.2-124M camera, besides standard optical components.

\subsection* {Code and Data Availability} 
The code developed for this study, along with a subset of the data, is openly available on GitHub at \href{github.com/ArcetriAdaptiveOptics/SPECULA}{github.com/ArcetriAdaptiveOptics/SPECULA} and \href{github.com/ArcetriAdaptiveOptics/SpecuLab}{github.com/ArcetriAdaptiveOptics/SpecuLab}. However, certain datasets, specifically the influence functions of the Deformable Mirror and related calibration data, are not publicly available, as they are an integral part of scientific projects. The corresponding author (fabio.rossi@inaf.it) can provide these data upon reasonable request and subject to approval of the relevant scientific collaborations.

\newpage

\section{Appendix A - Data and Processing classes}\label{sec:appendixA}

\begin{table}[h!]
\centering
\renewcommand{\arraystretch}{1.1}
\footnotesize
\resizebox{\textwidth}{!}{%
\begin{tabular}{|p{2.7cm}|p{4.5cm}|p{7.0cm}|}
\hline
\textbf{Category} & \textbf{Class Name} & \textbf{Description} \\
\hline

\multirow{3}{*}{Optical Wavefronts} 
& \texttt{ElectricField} & Complex amplitude and phase of a coherent wavefront. \\
& \texttt{Intensity} & Detected intensity maps (squared modulus of the electric field). \\
& \texttt{Pixels} & Digitized detector readouts (shot noise, ADC effects). \\
\hline

\multirow{4}{*}{Wavefront Sensing}
& \texttt{Slopes} & Wavefront sensor measurements (e.g. Shack-Hartmann: x, y centroid displacements). \\
& \texttt{PupData} & Geometry and validity maps of pixels in a Pyramid wavefront sensor. \\
& \texttt{SubapData} & Geometry and validity maps of subapertures in a Shack-Hartmann sensor. \\
& \texttt{Lenslet} & Properties of a Shack-Hartmann lenslet array. \\
\hline

\multirow{3}{*}{System Geometry}
& \texttt{PupilStop} & Pupil masks and obstruction patterns defining telescope pupil geometry. \\
& \texttt{Layer} & A single turbulence layer or optical phase screen (altitude, electric field). \\
& \texttt{Source} & Guide stars and science targets (position, magnitude, spectrum). \\
\hline

\multirow{4}{*}{Calibration Data}
& \texttt{Intmat} & Maps deformable mirror commands to wavefront sensor slopes. \\
& \texttt{Recmat} & Converts wavefront sensor slopes into deformable mirror commands. \\
& \texttt{IFunc} & Influence function of a deformable mirror actuator. \\
& \texttt{M2C} & Mode-to-command transformation matrix for modal control loops. \\
\hline

\multirow{4}{*}{Signal Processing}
& \scriptsize{\texttt{ConvolutionKernel}} & Generic container for convolution kernels. \\
& \scriptsize{\texttt{GaussianConvolutionKernel}} & Kernel generator for Gaussian kernels. \\
& \texttt{IirFilterData} & Coefficients for Infinite Impulse Response (IIR) filters. \\
& \texttt{TimeHistory} & Temporal sequences of data for filtering/analysis. \\
\hline

\multirow{2}{*}{Other Data}
& \texttt{LaserLaunchTelescope} & Geometry and parameters of a laser guide star launcher. \\
& \texttt{InfinitePhaseScreen} & Infinite turbulence phase screen for seamless simulation. \\
\hline
\end{tabular}}
\caption{Overview of data objects defined so far in SPECULA.}
\label{tab:dataObjects}
\end{table}

\newpage

\begin{table}[h!]
\centering
\renewcommand{\arraystretch}{1.1}
\footnotesize
\resizebox{\textwidth}{!}{%
\begin{tabular}{|p{1.6cm}|p{4.3cm}|p{8.3cm}|}
\hline
\textbf{Category} & \textbf{Class Names} & \textbf{Description} \\
\hline

\multirow{6}{*}{Generators}
& \texttt{WaveGenerator} & Generates different waveforms: \scriptsize{SIN, SQUARE, TRIANGLE.} \\
& \texttt{VibrationGenerator} & Generates vibration signals from PSD specifications. \\
& \texttt{RandomGenerator} & Generates random data. \\
& \texttt{PushPullGenerator} & Generates push-pull signals for modal calibration. \\
& \texttt{TimeHistoryGenerator} &  Generates signals from pre-computed time history data. \\
& \texttt{ScheduleGenerator} &  Generates scheduled values that change at specified times. \\
\hline

\multirow{3}{*}{Atmosphere}
& \texttt{AtmoEvolution} & Turbulence model based on finite phase screens. \\
& \texttt{AtmoInfiniteEvolution} & Turbulence model based on infinite phase screens. \\
& \texttt{AtmoRandomPhase} & Random phase of given spectral characteristics. \\
\hline

\multirow{2}{*}{Propagation}
& \texttt{AtmoPropagation} & Implements propagation using the Fourier model. \\
& \scriptsize{\texttt{ElectricFieldCombinator}} & Implements the sum of Electric Fields. \\
\hline

\multirow{4}{*}{WFS}
& \scriptsize{\texttt{ModulatedPyramid}, \texttt{ModulatedDoubleRoof}, \texttt{PolyCromPyramid}}  & Different kinds of Pyramid wavefront sensor. \\
& \scriptsize{\texttt{SH}, \texttt{DistributedSH}, \texttt{PolyChromSH}} & Different kind of Shack-Hartmann wavefront sensor. \\
& \texttt{ZernikeSensor} & Zernike wavefront sensor. \\
& \texttt{IdealDerivativeSensor} & Ideal derivative sensor. \\
\hline

\multirow{4}{*}{Other HW}
& \texttt{FocalPlaneFilter} & Focal Plane Filter. \\
& \texttt{CCD} & CCD detector model. \\
& \texttt{DeformableMirror} & Deformable mirror model. \\
& \texttt{PsfCoronagraph} & Perfect coronagraph implementation. \\
& \texttt{Psf} & Psf computation. \\
\hline

\multirow{7}{*}{Control}
& \scriptsize{\texttt{ShSlopec}, \texttt{PyrSlopec}, \texttt{DoubleRoofSlopec}} & Different kinds of slopes computers, for different WFSs. \\
& \texttt{IirFilter} & Infinite Impulse Response filter based Time Control. \\
& \texttt{OpticalGainEstimator} & Optical Gain Estimator, based on demodulated signals. \\
& \texttt{GainOptimizer} & Modal Gain optimizer for IIR filters. \\
& \texttt{Demodulator} & Extracts signals for specific control strategies. \\
&  \scriptsize{\texttt{Modalrec}, \texttt{ModalrecImplicitPolc}} & Different kind of reconstructors, compute deformable mirror commands from estimated wavefront. \\
& \texttt{WindowedIntegration} & Simple windowed integration of a signal. \\
\hline

\multirow{4}{*}{Calibrators}
& \texttt{ImCalibrator} & Calibrator to obtain the interaction matrix. \\
& \texttt{RecCalibrator} & Calibrator to obtain reconstructors. \\
& \texttt{PyrPupdataCalibrator} & Calibrator to obtain pupil maps for pyramid WFS. \\
& \texttt{ShSubapCalibrator} & Calibrator to obtain subapertures maps for SH WFS. \\
\hline

\multirow{4}{*}{Other}
& \texttt{MathOp} & Performs specific mathematical operations. \\
& \texttt{ConcatVectors} & Concatenates vectors or matrices. \\
& \texttt{PhaseFlattening} & Removes the mean phase from an electric field. \\
& \texttt{ExtendedSource} & Extended sources for pyramid wavefront sensing. \\
\hline

\multirow{3}{*}{Data Mgmt}
& \texttt{DataStore} & Saves time series of a set of Processing Objects outputs. \\
& \texttt{DataLoad} & Reloads saved time series. \\
& \texttt{DisplayServer} & Implements a Flask web server, see 
Sect.\ref{sec:visual}. \\
\hline

\end{tabular}}
\caption{Overview of Processing Objects implemented so far in SPECULA.}
\label{tab:procObjects}
\end{table}

\newpage

\section{Appendix B - Configuration files details, syntax and examples}\label{sec:appendixB}

\vspace{1ex}

The file usually begins with a main section, which corresponds to a data object of the class \texttt{SimulParams}, containing several general parameters regarding the simulation, like the folder used as root directory for the data storage or the duration in seconds of a time step.
The definitions of all the data and processing objects in the simulation are as follows. Note that the data objects directly specified in this direct way are simply constant parameters which are needed by some processing object.

\begin{listing}[h!]
\begin{lstlisting}[language=yaml, gobble=4]
    object_nameB:
      class: 'ObjectClass'
      scalar_param: 23
      namedDataObject_object: 'filename1'
      namedRawData_data: 'filename2'
      singleDataObject_ref: 'object_name0'
      listOfDataObjects_dict_ref: ['object_name1', 'object_name2', ...]
      inputs:
        in_i: 'object_nameA.out_name'
      outputs:  ['out_name1', 'out_name2', ...]
\end{lstlisting}
\caption{YAML processing object definition.}
\label{listTxt:1}
\end{listing}

A template for a generic processing object section is shown in Listing \ref{listTxt:1}.
After specifying the name of the object and its type (class), the signature of the constructor of its class must be matched, specifying all of its parameters values, when different from the default ones. In our example \texttt{scalar\_param} is set to 23. Some parameters might be raw data arrays that need to be loaded from disc, and in this case the postfix \texttt{\_data} is added to the name of the parameter that appears in the constructor, while the corresponding value is a string containing the file name to load (without the \texttt{.fits} extension). Some parameters might be data objects that need to be restored from disc, and in this case the postfix \texttt{\_object} is added to the name of the parameter that appears in the constructor, while the corresponding value is a string containing the file name to load.
Finally, some of the parameters can be data objects which are specified somewhere in the configuration files: in this case the postfix \texttt{\_ref} is added to the parameter name, and the value is the name of the referred data object (\texttt{\_dict\_ref} is used in the case of a list of objects). The parameters  having a default value in the python class specification of the processing object might be omitted if the default value is to be used.
After this specification of the class constructor parameters, the inputs and outputs are specified: care must be taken to match the actual inputs and outputs defined in the processing object code. The inputs are a sequence of {key: value} pairs, with input names as keys and references to some other object output as values, in the format: 
\begin{lstlisting}[language=yaml, gobble=0, frame=none]
input_name: 'object_name.output_name'
\end{lstlisting}

The outputs are just a list of output names. Note how some processing object might not have any inputs.
The sections describing data objects have the same format as the ones for processing objects, except that they have neither inputs nor outputs. 
It is left to the user to connect inputs and outputs of the same data object class: failing this requirement will produce an error message during the simulation initialization.

The syntax to specify the incremental modifications to the initial specification file is the following:

\begin{itemize}
    \item To add a new section/object, just write the section defining the new object
    \item To modify a specific parameter:
\begin{lstlisting}[language=yaml, gobble=0, frame=none]
    objName_override: { param_key: new_param_value }
\end{lstlisting}
    \item To remove one or more objects:
    write a list of the objects to be removed: 
\begin{lstlisting}[language=yaml, gobble=0, frame=none]
    remove: ['objToRemove1', 'objToRemove2',  ...]
\end{lstlisting}
\end{itemize}

The format to specify \emph{DataStore} inputs is:
\begin{lstlisting}[language=yaml, gobble=0, frame=none]
inputs:
  input_list: ['aFilename-processingObjectNameA.outputNameA', 
    'bFilename-processingObjectNameB.outputNameB', ... ]
\end{lstlisting}

\begin{spacing}{1}
\begin{lstlisting}[language=yaml, basicstyle=\tiny\ttfamily, gobble=0, frame=single, numbers=left]
main:
  class:             'SimulParams'
  root_dir:          '/raid/LBT'              # Root directory for calibration manager  
  pixel_pupil:       220                      # Linear dimension of pupil phase array
  pixel_pitch:       0.037372727              # [m] Pitch of the pupil phase array
  total_time:        10.9956                  # [s] Total simulation running time
  time_step:         0.000588                 # [s] Simulation time step
  zenithAngleInDeg:  0.000
  display_server:    False

seeing:
  class:             'WaveGenerator'
  constant:          0.8                      #  seeing value

wind_speed:
  class:             'WaveGenerator'
  constant:          [2.,4.,6.,25.]           # [m/s] Wind speed value

wind_direction:
  class:             'WaveGenerator'
  constant:          [0.,30.,-30.,45.]        # [degrees] Wind direction value
  
on_axis_source:
  class:             'Source'
  polar_coordinates: [0.0, 0.0]               # [arcsec, degrees] source polar coordinates
  magnitude:         8                        # source magnitude
  wavelengthInNm:    750                      # [nm] wavelength

pupilstop:                                    # Default parameters (circular pupil)
  class: 'Pupilstop'
  simul_params_ref: 'main'

atmo:
  class:                 'AtmoEvolution'
  simul_params_ref:      'main'
  L0:                    40                    # [m] Outer scale
  heights:               [119.,837,3045,12780] # [m] layer heights at 0 zenith angle
  Cn2:                   [0.70,0.06,0.14,0.10] # Cn2 weights (total must be eq 1)
  fov:                   240.0
  seed:                  1
  inputs:
    seeing: 'seeing.output'
    wind_speed: 'wind_speed.output'
    wind_direction: 'wind_direction.output'
  outputs: ['layer_list']

prop:
  class:                 'AtmoPropagation'
  simul_params_ref:      'main'
  source_dict_ref:       ['on_axis_source']
  inputs:
    atmo_layer_list: ['atmo.layer_list']
    common_layer_list: ['pupilstop',
                  'dm.out_layer:-1']
  outputs: ['out_on_axis_source_ef']

modal_analysis:
  class:                 'ModalAnalysis'
  ifunc_inv_object:      'KLmatrix_KL_v10_inv'
  inputs:
    in_ef: 'prop.out_on_axis_source_ef'
  outputs: ['out_modes']

pyramid:
  class:             'ModulatedPyramid'
  simul_params_ref:  'main'
  pup_diam:          40.                      # Pupil diameter in subaps.
  pup_dist:          48.                      # Separation between pupil centers in subaps.
  fov:               2.0                      # Requested field-of-view [arcsec]
  mod_amp:           3.0                      # Modulation radius (in lambda/D units)
  output_resolution: 120                      # Output sampling [usually corresponding to CCD pixels]
  wavelengthInNm:    750                      # [nm] Pyramid wavelength
  inputs:
    in_ef: 'prop.out_on_axis_source_ef'

detector:
  class:             'CCD'
  simul_params_ref:  'main'
  size:              [120,120]                # Detector size in pixels
  dt:                0.000588                 # [s] Detector integration time
  bandw:             300                      # [nm] Sensor bandwidth
  photon_noise:      False                    # activate photon noise
  readout_noise:     False                    # activate readout noise
  excess_noise:      False
  readout_level:     0.4                      # readout noise in [e-/pix/frame]
  quantum_eff:       0.32                     # quantum efficiency * total transmission
  inputs:
    in_i: 'pyramid.out_i'

slopec:
  class:             'PyrSlopec'  
  # tag of the pyramid WFS pupils
  pupdata_object: 'soul_KLv10_ps220p0.037_pyr40x40_wl750_fv2.1_ft3.0_bn1_th0.30a0.30b'            
  inputs:
    in_pixels:         'detector.out_pixels'
  outputs: ['out_slopes']

rec:
  class:               'Modalrec'
  # reconstruction matrix tag
  recmat_object:       'soul_KLv10_ps220p0.037_pyr40x40_wl750_fv2.1_ft3.0_ma3_bn1_th0.30a0.30b_mn500'
  inputs:
    in_slopes:         'slopec.out_slopes'
  outputs: ['out_modes']

control:
  class:             'Integrator'
  simul_params_ref:  'main'
  delay:             2                        # Total temporal delay in time steps
  int_gain:          [0.55,   0.45,     0.40,     0.35]
  ff:                [1.0,    0.999996, 0.999365, 0.977895]
  n_modes:           [30,     70,       100,      300]
  inputs:
      delta_comm: 'rec.out_modes'

dm:
  class:             'DM'
  simul_params_ref:  'main'
  ifunc_object:      'KLmatrix_KL_v10'
  height:            0                        # DM height [m]
  inputs:
      in_command: 'control.out_comm'

psf:
  class:             'PSF'
  simul_params_ref:  'main'
  wavelengthInNm:    1650                     # [nm] Imaging wavelength
  nd:                8                        # padding coefficient for PSF computation
  start_time:        0.05                     # PSF integration start time
  inputs:
      in_ef:  'prop.out_on_axis_source_ef'

data_store:
  class:             'DataStore'
  store_dir:         './output'               # Data result directory: 'store_dir'/TN/
  data_format:       'fits'
  inputs:
    input_list: ['wfsPhi-prop.out_on_axis_source_ef',
                 'resMod-modal_analysis.out_modes',
                 'ccdframes-detector.out_pixels',
                 'slopes-slopec.out_slopes',
                 'deltaComm-rec.out_modes',
                 'comm-control.out_comm',
                 'srRes-psf.out_sr']
\end{lstlisting}
\end{spacing}
\vspace*{3mm}
\captionof{listing}{Full simulation definition file for the SOUL system.}
\label{listTxt:soulConfig}

\begin{listing}[h!]
\begin{lstlisting}[language=bash, basicstyle=\scriptsize\ttfamily, gobble=0, frame=single, numbers=left, breaklines=true]
#!/bin/bash

#SBATCH --job-name=spMSPREAD              # Descriptive job name
#SBATCH --time=00:10:00                   # Maximum wall time (hh:mm:ss)
#SBATCH --nodes=4                         # Number of nodes to use
#SBATCH --ntasks-per-node=4               # Number of MPI tasks per node (e.g., 1 per GPU)
#SBATCH --cpus-per-task=1                 # Number of CPU cores per task (adjust as needed)
#SBATCH --gres=gpu:4                      # Number of GPUs per node (adjust to match hardware)

#SBATCH --partition=boost_usr_prod        # GPU-enabled partition
#SBATCH --qos=normal                      # Quality of Service
#SBATCH --output=jobMorfeoMpiSpread.out   # File for standard output
#SBATCH --error=jobMorfeoMpiSpread.err    # File for standard error
#SBATCH --account=try25_rossi             # Project account number

module load cuda/12.3                     # Load CUDA toolkit
module load openmpi/4.1.6--nvhpc--23.11   # Load MPI implementation
module load nvhpc/23.11

srun --mpi=none bash -c "specula --mpi $WORK/SPECULA/config/MORFEO/params_morfeo_full.yml $WORK/SPECULA/config/leonardo/overrides/ov_MORFEO_leonardo_spread.yml"
\end{lstlisting}
\caption{Configuration file for slurm to run over 4 computing nodes (16 GPUs).}
\label{listTxt:slurm}
\end{listing}

\newpage

\bibliography{report}   
\bibliographystyle{spiejour}   


\vspace{2ex}\noindent\textbf{Fabio Rossi} Fabio Rossi is a software engineer and PhD in Computer Science. For almost a decade, he has been working at INAF, Arcetri Astrophysical Observatory, as a member of the Adaptive Optics group. He was involved in the design, development and commissioning of SOUL, AO system for LBT, and he is currently involved in ELT ANDES and VLT MAVIS. His research interests include Adaptive Optics simulation and software design, High Performance Computing, and Machine Learning. He is one of the main developers of the analytical AO simulation framework TipTop.

\vspace{2ex}\noindent\textbf{Guido Agapito} Guido Agapito is a senior researcher at INAF, Arcetri Astrophysical Observatory, with more than 15 years of experience in AO for astronomy. He was involved in the design and commissioning of AO instruments such as VLT ERIS and LBT SOUL and he is currently involved in ELT MORFEO and VLT MAVIS. He is an expert in AO control systems, the main developer of the PASSATA and a developer of the analytical AO simulation framework TipTop.

\vspace{2ex}\noindent\textbf{Alfio Puglisi} Alfio Puglisi is a software engineer at INAF, Arcetri Astrophysical Observatory. He has more than 20 years of experience in Adaptive Optics, ranging from laboratory systems to instrument design, integration and commissioning. He contributed most of the GPU acceleration code to PASSATA and is currently working on the MORFEO real-time computer for the ELT.

\vspace{1ex}

\listoffigures
\listoftables

\end{spacing}
\end{document}